\documentclass[%
 reprint,
nofootinbib,
 amsmath,amssymb,
 aps,
 twocolumn,
]{revtex4-2}

\usepackage{graphicx}
\usepackage{dcolumn}
\usepackage{bm}
\usepackage{hyperref}
\usepackage{color}
\usepackage{enumitem}

\usepackage{booktabs, multirow} 
\usepackage{soul}
\usepackage{xcolor,colortbl} 
\usepackage{changepage,threeparttable} 
\usepackage{tcolorbox}

\begin{document}


\title{Pulses, waves, and cascades in collective migration dynamics}

\author{Niraj Kushwaha$^{1,2,3}$}%
\author{Woi Sok Oh$^3$}%
\author{Edward D. Lee$^{2,4}$}
 \email{edlee@csh.ac.at}
\affiliation{%
 $^1$Faculty of Physics, University of Vienna, Boltzmanngasse 5, Vienna, 1090, Vienna, Austria\\
 $^2$Complexity Science Hub, Metternichgasse 8, 1030 Vienna, Austria\\
 $^3$Department of Systems Design Engineering, University of Waterloo, 200 University Ave W, Waterloo, ON N2L 3G1, Canada\\
 $^4$Institute of Forest Ecology, University for Natural Resources and Life Sciences, Peter-Jordan-Straße 82, 1190 Vienna, Austria
}%

\date{\today}

\begin{abstract}
Decisions to migrate depend on others' decisions. Dependence can produce nontrivial dynamics. We propose a minimal migration model that accounts for social influence alongside individual heterogeneity in mobility as migrants move from region to region. In special locations of parameter space, migrant flows dramatically and spontaneously fluctuate. Such aspects mimic observed fluctuations in migration statistics and thus show how large fluctuations in data can reflect more than response to events like armed conflict and natural disasters. Correspondingly, the impact of exogenous factors can be confounded with the results of collective decisions.
\end{abstract}


\maketitle

\begin{figure}
\centering
	\includegraphics[width=\linewidth]{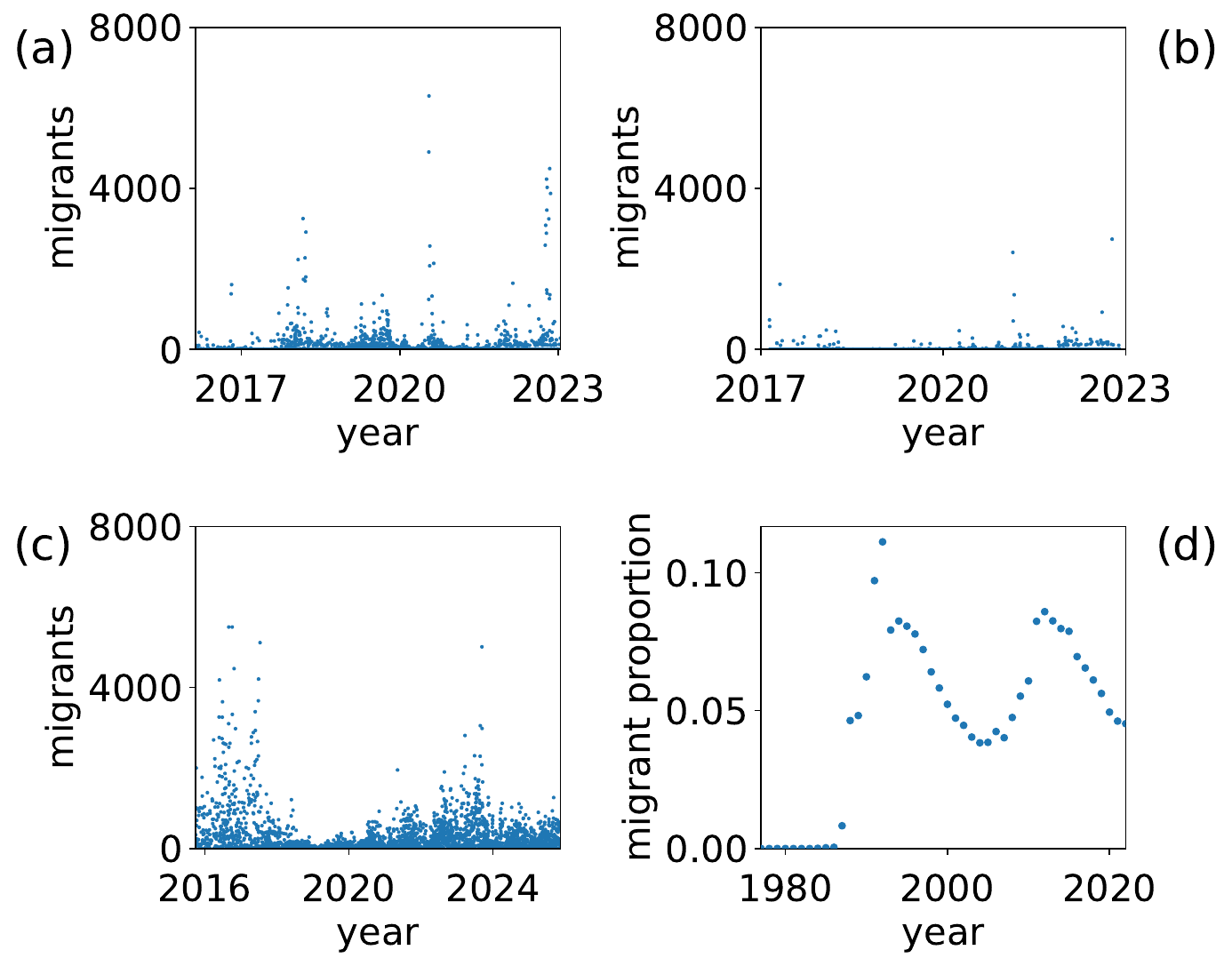}
	\caption{Fluctuations in observed migration rates. (a) Daily number of surveyed migrants heading into Banadir from Afgooye or (b) from WanlaWeyn into Banadir, all districts in Somalia \cite{unhcrPRMNDashboard2022}. (c) Daily number of maritime migrants landing in Sicily \cite{unhcrSituationEurope2026}. Standard deviation $\sigma=515$ and mean $\mu=348$. (d) Estimated annual refugee and asylum seekers as a proportion of all outgoing emigrants originating in Somalia~\cite{susmannBayesianProjection2025}.}\label{fig:data}
\end{figure}

Suddenly changing migration patterns have historically been a major social and political concern \cite{shanahanEffectsImmigrant1999, schragNotFit2010}, and abrupt, collective changes are characteristic of historical migration data. For example,  fine-grained data of Somali internal migration reveal that the daily number of migrants $n$ through checkpoints exhibits distributions with tails close to power law, $p(x)\propto x^{-\alpha-1}$ for $\alpha>1$, or multi peaked distributions (Figure~\ref{fig:data}). Daily maritime arrivals landing in Sicily show fluctuations, with noise ratio of standard deviation to mean exceeding unity, $\eta=3/2$. Over longer time scales of decades, outgoing refugee and asylum-seeker flows from countries change suddenly and dramatically: some estimates show relative increase of 10 times in 5 years \cite{susmannBayesianProjection2025}. These examples prominently feature fluctuations \cite{gaskinDeepLearning2025}. Yet, assumptions of steady state have reigned in quantitative models of migration. 

Stability is, for example, a central assumption in the push-and-pull framework \cite{lee1966theory}, which stipulates that migration between two locations consists of two factors: one for why people leave the origin (e.g.,~economic hardship) and the other for which someone heads for a particular destination (e.g.,~earnings, good weather, close relations \cite{sjaastadCostsReturns1962}). While these factors can themselves change suddenly, widely used examples of the general framework like the gravity \cite{andersonGravityModel2011, parkGeneralizedGravity2018} and radiation models \cite{siminiUniversalModel2012}, the workhorses of the migration literature, do not alone account for such abrupt changes \cite{beyerGravityModels2022}. 
When fluctuations are addressed, they are often treated as random sources of statistical variability \cite{zensDynamicCount2025} or 
attributed to exogenous drivers \cite{cohenInternationalMigration2008, cappelenForecastingImmigration2015, suleimenovaGeneralizedSimulation2017, qiModellingPredicting2023, ghorbaniFlee32024}, not least demography \cite{plane1993demographic,foster2017decomposing}. In other work, they are fit as curves with a phenomenological and data-based, if accurate, model \cite{susmannBayesianProjection2025} without consideration of their origins in collective decisions \cite{bircanEditorialNew2025}. In addition, agent-based models could in principle incorporate a micro-level explanation of collective fluctuations, but they usually reflect assumptions of stability \cite{reichlovaCanMotivation2007}. Thus, migration models do not treat large fluctuations in the choice to migrate as central but rather as corrections to the underlying and implicitly steady-state dynamics.

From a statistical physics perspective, large fluctuations are particularly interesting because they emerge from interactions between components \cite{castellanoStatisticalPhysics2009, bialekSocialInteractions2014, leeStatisticalMechanics2015}. These fluctuations are not fully captured when incorporated as external drivers or random noise to steady-state dynamics. From this point of view, stability embedded in migration model structure is misaligned with the collective nature of migration decision-making. Such decisions depend on others' decisions, as in bandwagon effects or chain migration \cite{macdonaldChainMigration1964}, leading to large fluctuations in migration rates \cite{couzinEffectiveLeadership2005, couzinUninformedIndividuals2011}. 
The distribution of fluctuations, the emergence of scaling laws, or the presence of discontinuities are signatures of phase transitions emerging from local interactions. 
Since origin and destination are coupled through migrant flow, such phase transitions, if they exist, would be coupled dynamically across regions. This raises the possibility of nontrivial, endogenous dynamics leading to pulses, waves, and cascades. It is thus surprising that aside from exogenous forcing \cite{hoffmannMetaanalysisCountrylevel2020, thalheimerLargeWeather2023}, nonlinear dynamics are often neglected in both local or global models of migration.

Here, we consider migration as a stochastic nonlinear dynamical system to demonstrate the emergence of instability, recurrence, and cascades from interdependent decisions to migrate. We propose a simple model incorporating central tenets of migration. First, individuals cannot always or do not wish to migrate, so either they are capable of migration, ``mobile,'' or are incapable, ``immobile'' \cite{schewelUnderstandingImmobility2020, czaikaMigrationDrivers2022}. This is a distinction between individuals who have the wherewithal and desire to move versus those who do not. Second, individuals have different propensities for mobility \cite{carlingMigrationAge2002, czaikaMigrationDecisionMaking2021}. This heterogeneity in the population could reflect variation in abilities, resources, or aspirations to move. Third, the mobility of an individual encourages the mobility of others \cite{macdonaldChainMigration1964, epsteinInformationalCascades2002}. This is the key element of social influence. We describe a simple model that incorporates the three tenets, and we identify the regimes of dynamical stability and instability. 

Parts of parameter space have steady-state dynamics, fixed rates of migration between regions, and small fluctuations; these are similar to the assumptions underlying the push-and-pull framework. More surprisingly, we find special parameter values that display either spatial or temporal symmetry breaking with abrupt collective transitions in mobility. In the former, dynamical instabilities grow to exaggerate differences in population density, providing an original explanation for how collective decisions can lead to variable populations. In the latter, dynamics are characterized by large fluctuations with recurrent, periodic episodes of high and low migration. These are preceded by random, intermittent fluctuations in migration rates. Intermittent fluctuations can lead to cascading changes in migration rates across multiple regions, depending on migration route network connectivity. Finally, we consider how prolonged exogenous shocks that force inhabitants out of regions can lead to long-term changes in migration patterns that are ``remembered'' in the local population sizes and dynamics. This set of dynamical outcomes underlines the nontrivial role of social influence in collective migration. If such factors are left out of models, then endogenous dynamics can be confounded with potential drivers such as political violence and natural disasters \cite{bijakAssessingUncertain2020, thalheimerLargeWeather2023}.

\section{Collective migrant-flow model}
\begin{figure}
	\centering
	\includegraphics[width=\linewidth]{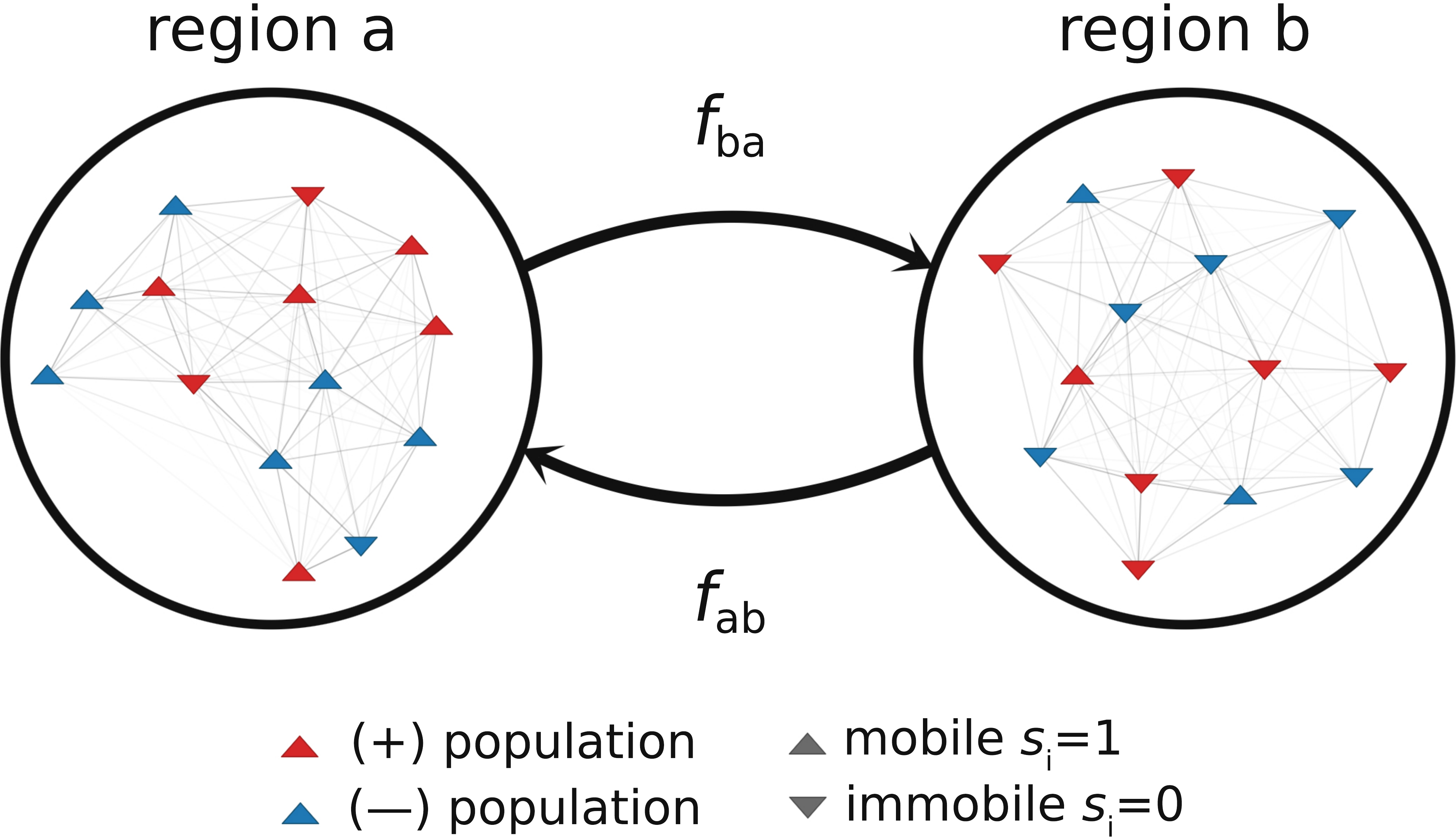}
	\caption{Model diagram. Two regions labeled a and b are coupled via migrant flow from one to the other. Within each region, each individual belongs to either the (blue) sedentary ($-$) or (red) more mobile ($+$) population. Each individual i is either (up triangle) mobile $s_{\rm i}=1$ or (down triangle) immobile $s_{\rm i}=0$ with probability that depends on individual bias and interactions.}
	\label{fig:model diagram}
\end{figure}

To formulate a model that accounts for the three tenets (mobility vs.~immobility, population heterogeneity, social influence), we propose a general and yet minimal way of accounting for each individual's preferences. Each migrant i can be either in a mobile state $s_{\rm i}=1$ or an immobile state $s_{\rm i}=0$. When a migrant is mobile, a migrant in region a moves with rate $f_{\rm ba}$ to region b as we show in Figure~\ref{fig:model diagram}a. When migrants are immobile, they do not move.

To account for overall propensity to be found in either configuration, we would like to account for each individual's intrinsic propensity in addition to social influence. A minimal preference function for individual i would then be 
\begin{align}
	h^{\rm eff}_{\rm i}(t) &= h_{\rm i} + J m_{\rm a}(t),\label{eq:heff}
\intertext{where we translate Eq~\eqref{eq:heff} into a probability of being mobile}
	p(s_{\rm i}=1;t) &= \frac{e^{h^{\rm eff}_{\rm i}(t)}}{1 + e^{h^{\rm eff}_{\rm i}(t)}}.\label{eq:p(si)}
\end{align}
Taken together with Eq~\eqref{eq:p(si)}, $h^{\rm eff}_{\rm i}(t)$ constitutes a time $t$ dependent ``effective field'' that indicates a bias towards mobility when positive and towards immobility when negative. Eq~\eqref{eq:heff} contains a bias $h_{\rm i}$ that represents a personal tendency for mobility (again, when positive it heightens the typical mobility) and a ``coupling'' $J>0$, which specifies the strength of influence from the typical mobility of others.\footnote{The coupling in the Hamiltonian of a region is normalized by population size to model the fact that social ties and social dependence increases as population size decreases \cite{stickImmigrantsSocial2024, greiderNeighboringPatterns1985}.} 
The typical {\it mobility} of others is encapsulated in the time-dependent average in region a, or ``magnetization'' in statistical physics,
\begin{align}
	m_{\rm a}(t) &= \frac{1}{n_{\rm a}(t)} \sum_{\rm i=1}^{n_{\rm a}(t)} s_{\rm i}(t)
\end{align}
normalized by the total population $n_{\rm a}(t) \equiv n^+_{\rm a}(t)+n^-_{\rm a}(t)$, a time-dependent quantity since migrants move from one region to another. 
Eqs~\eqref{eq:heff} and \eqref{eq:p(si)} together constitute the familiar mean-field Ising model from statistical physics, a minimal model of interactions \cite{jaynesInformationTheory1957}. Thus, they summarize how each migrant has an inherent tendency to be either mobile or immobile, but the coupling means that the overall tendency to be mobile increases with the addition of every other mobile migrant. 

To account for individual-level variability, we distinguish two migrant populations with superscripts ($+$) and ($-$), each numbering $n^+_{\rm a}(t)$ and $n^-_{\rm a}(t)$ with the former having a higher tendency for mobility and the latter a lower tendency for mobility. 
More formally, this corresponds to a binomial distribution for the individual biases, or $p(h_{\rm i}=h^+)=q$ and $p(h_{\rm i}=h^-)=1-q$, where $h^+> h^-$. For each respective population, we must reformulate the mean-field proposal for the effective field from Eq~\eqref{eq:heff} to consider the impact of the mixed population and obtain now the two effective fields $\tilde{h}_{\rm a}^{\pm} \equiv h^\pm + Jm_{\rm a}(t) - Jm_{\rm a}^\pm(t)/n_{\rm a}$,\footnote{For consistency when regions reach a size of 1, we include the correction for self-interaction.} where the time-dependent magnetizations of the respective populations are $m_{\rm a}^\pm(t) = \sum_{\rm i\in\pm}s_{\rm i}/n_{\rm a}^\pm(t)$ and the magnetization of the entire population the weighted average 
\begin{align}
    m_{\rm a} = (n_{\rm a}^+m_{\rm a}^+ + n_{\rm a}^-m_{\rm a}^-)/n_{\rm a}.
\end{align}
Thus, the mean-field theory for the binomial random-field Ising model (RFIM) obeys a form similar to the usual mean-field theory as in Eq~\eqref{eq:heff}, but the dynamical coupling between the populations introduces co-dependence in their respective mobilities.

The explicit dependence on time accounts for the rates at which mobile migrants move from one region to another and how quickly they are persuaded to mobility by others. In the directionally symmetric version of the route network, the rate of migration is symmetric and constant $f_{\rm ab}=f_{\rm ba}=f$, or that the total rate of outmigration from region a is $fn_{\rm a}(t)m_{\rm a}(t)$.\footnote{Heterogeneity in $f$, still assuming symmetric rates $f_{\rm ab}=f_{\rm ba}$, should only affect the steady-state configurations but not the kinds of dynamics that we observe. Since we are only interested in the latter, we take a uniform $f$.} The second relevant rate is how quickly the migrants at each region adjust their mobility based on the local milieu over an equilibrium time $\tau$. When $f\gg1/\tau$, migrants flit from one region to another quickly, only weakly interacting with the local population, and in the opposing limit $f\ll1/\tau$ the migrants equilibrate first and foremost. The former takes us to the limit of a fully-connected mean-field Ising model for all populations, erases the spatial structure given by the regions, and is thus uninteresting, whereas the latter limit is the one where migrant flows can lead to symmetry breaking between the regions.

Then, taking the assumption that equilibration happens much faster than migration, each region will satisfy the two self-consistency conditions,\footnote{We solve Eq~\eqref{eq:self-consistency} numerically by finding candidate roots with scipy's fsolve routine and then refining the resulting solutions with an implementation of Halley's method, which allows us to obtain the solutions with high precision.}
\begin{align}
\begin{aligned}
	m_{\rm a}^\pm(t) &= \frac{e^{\tilde{h}_{\rm a}^\pm(t)}}{1 + e^{\tilde{h}_{\rm a}^\pm(t)}}.
\end{aligned}\label{eq:self-consistency}
\end{align}
Besides the regions of the stable solution manifold that are continuous (as in the paramagnetic regime), the solution manifold can fold over itself, displaying bifurcations and the emergence of discontinuous jumps as a function of the interaction term $J m$, which itself depends on a region's populations $n^+$ and $n^-$ as we show in Figure~\ref{fig:solution landscape}. In some cases, the binomial RFIM can have up to three stable solutions when fold bifurcations in the two populations interact. We will need to account for this nontrivial structure when we consider the dynamical outcomes.

We account for the stochastic flow of migrants from one region to another, again starting with the assumption that migrants equilibrate quickly in their respective regions. We write down the stochastic dynamics that account for migrant flow out of region a and from all neighboring regions b for the ($+$) and ($-$) populations, respectively,
\begin{align}
	dn_{\rm a}^{\pm}(t) &= - \sum_{\rm b \neq a}\left[ d\mathcal{N}_{\rm a}^{\pm}(t) -  d\mathcal{N}_{\rm b}^{\pm}(t)\right]. \label{eq:dn stochastic}\\
\intertext{The random variables have expectations}
	\mathbb{E}\!\left[d\mathcal{N}_{\rm a}^{\pm}(t)\right] &= \frac{f}{d_{\rm a}^{\rm out}}\,n_{\rm a}^{\pm}(t)\,m_{\rm a}^{\pm}(t)\;dt\\
	\mathbb{E}\!\left[ dn_{\rm a}^\pm\right] &= \left[-f\,n_{\rm a}^\pm m_{\rm a}^\pm + f\sum_{\rm b\neq a} A_{\rm ab} \frac{n_{\rm b}^\pm m_{\rm b}^\pm}{d_{\rm b}^{\mathrm{out}}}\right]dt \label{eq:dn mft}
\end{align}
for adjacency matrix $A_{\rm ab}$, when region b can send migrants to region a, and the total outdegree of region b $d_{\rm b}^{\rm out}\leq r-1$, given $r$ regions.  
We take migrant flow to be uncorrelated in time such that the distribution of migrants moving in a chosen small time interval is Poissonian. For the simulations we discuss here, we take $dt=1$. The implicit nonlinearities in Eq~\eqref{eq:dn stochastic}, given that the mobilities depend on the populations, hint at the possibility of nontrivial, long-term dynamics. When stochasticity drives the system across discontinuities in the solution manifold, they may lead to intermittent instability and recurrence.

Overall, the minimal migration model that we present here accounts for the collective nature of migration decisions, individual heterogeneity, and the randomness of migration flow between origin and destination. We approach this problem with both the mean-field formulation to permit some analytic transparency, but we also compare this calculation with a stochastic automaton simulation that accounts for the noisy process of decision making (further discussed in Section~\ref{sec:automaton}). As we show in the next sections, the relative strengths of such factors determine the emergence of nontrivial dynamics, sudden transitions, and regular excitations in migration, explaining how dynamical, collective instability arises endogenously.

\section{Symmetry breaking and intermittent pulses in two regions}\label{sec:pulses}
To gain intuition about the dynamics of the migrant flow model, we consider the simplest example of two regions a and b coupled to one another via migrant flows with symmetric rates $f_{\rm ab}=f_{\rm ba}=f$ (Figure~\ref{fig:model diagram}). There are a few general observations that we can make, and these are topological consequences of the solution manifold, which displays either a smooth or a sudden transition from low to high mobility as populations migrate.

\begin{figure}
\centering
	\includegraphics[width=\linewidth]{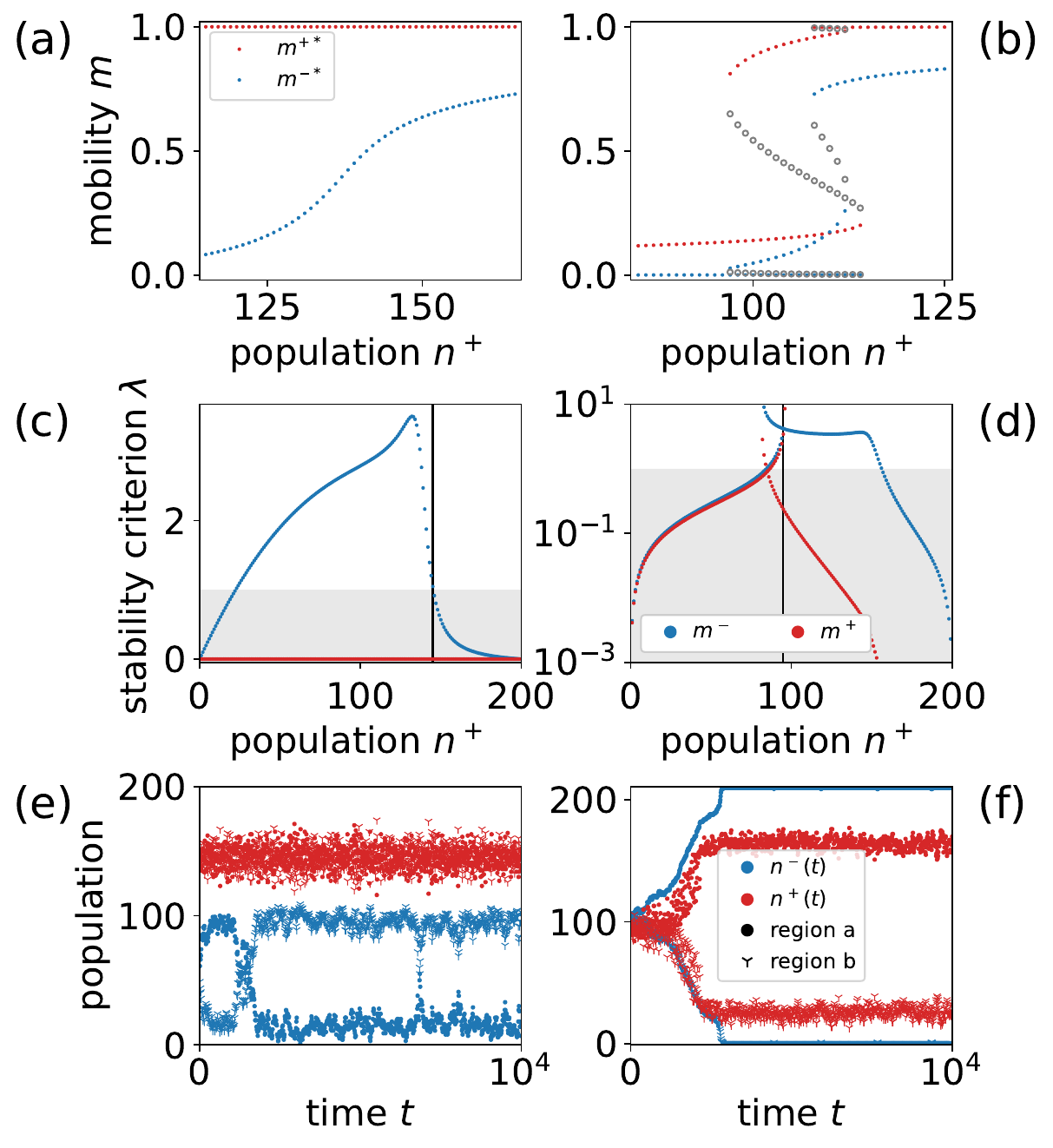}
\caption{(a) Example of a continuous mobility solution manifold as a function of ($+$) population $n^+$ with constant total population $n^++n^-=200$. (b) Example of an overlapping fold bifurcation as a function of ($+$) population $n^+$ again with constant total population. (c, d) Comparison of the two stability criteria colored red and blue, respectively. Vertical black line indicates where the dynamical simulations were run in panels c and d. Gray zone indicates stability. (e, f) Regional symmetry breaking under dynamical simulation in populations as predicted by linear stability analysis. (e) Close to marginal stability, the population $n^-$ shows large fluctuations.}
\label{fig:stability}
\end{figure}

Consider first the continuous transition from low to high mobility for both $m^+$ and $m^-$ as we increase either $n^+$ or $n^-$, respectively, as in Figure~\ref{fig:stability}a. This is synonymous with the paramagnetic phase.  
To determine dynamical outcomes, we ask how a small deviation $\delta n^\pm$ from the fixed-point solution $n_{\rm a}^\pm=n_{\rm b}^\pm = n^\pm/2$ grows, again indicating two separate terms for the ($+$) and ($-$) populations. In the linear regime, we find two contributions to its time evolution $\delta \dot n^\pm = -2fm^\pm - fn^\pm {m^\pm}'$. The first term indicates that deviations tend to self-dampen in proportion to the mobility.  The sign of the second term, however, depends on the response of the mobility to perturbation, given by the total derivative ${m^\pm}'$, which can be negative.  As we show in Appendix~\ref{sec:C_transition}, the conditions for instability are then
\begin{align}
	\lambda \equiv \left| \frac{d \log m^\pm}{d\log n^\pm} \right| > 1
\end{align}
We show examples of the stability criterion for the ($+$) and ($-$) populations in Figure~\ref{fig:stability} panels c and d, and we show how the dynamical simulations align with the stability criterion.

When the stability criterion is violated, we find symmetry breaking when an initial stochastic break of region-to-region population symmetry grows. Consider the case of ($-$) migrants: a decrease in $n^-$ increases mobility because the interaction term $Jm$ in Eq~\eqref{eq:heff} increases.\footnote{Note that $m$ is a weighted average of the two populations and by construction $m^+>m^-$, a decrease in $n^-$ or increase in $n^+$ increases the mobility $m^-$. More formally, its partial derivative with respect to $n^-$ is negative, $\partial m/\partial n^- = (m^--m)/n<0$.} When this is sufficiently strong to overcome the self-dampening flow, this picture implies that ($-$) migrants tend to ``evaporate,'' becoming more mobile the fewer of them there are---and inversely becoming less mobile the more of them there are---leading to regional disparities as they amass in some regions and disperse from others. For the two-region example, such dynamics eventually stabilize at distinct population numbers, whose ultimate difference is given by the curvature of the mobility response curve.

In contrast, we expect a different kind of intuition to apply to the ($+$) population because the partial derivative of the mobility $\partial m^+/\partial n^-$ has the opposite sign of $\partial m^-/\partial n^-$. With a continuous transition and conserved total population $n$, the ($+$) population always satisfies the stability criterion: a more populous region will have more mobile migrants. This increases migrant flow out of the more populous region and decreases mobility in the less populous region, therefore incurring an equalizing flow. 
The argument for the continuous manifold, formalized in Appendix \ref{sec:C_transition}, indicates that the long-term densities of the ($+$) and ($-$) populations can be characteristically different because flows of ($-$) migrants tend to break ``spatial'' symmetry between regions and flows of ($+$) migrants tend to balance out regional disparities.

A discontinuous transition in the mobility solution manifold leads to different outcomes than from the continuous solution manifolds. The ($+$) population can become dynamically unstable when the manifold has multiple branches as we show in the example of Figure~\ref{fig:stability}d. As a result, regional populations segregate into distinct mobilities as they come to occupy different branches of the solution manifold. By the previous argument for the continuous transition (in reference to the positive slope), this cannot occur alone for the ($+$) population but only when such symmetry breaking is driven by simultaneous changes in the ($-$) population such as when the fold bifurcations overlap (Figure~\ref{fig:stability}b).\footnote{Since the self-consistency conditions in Eq~\eqref{eq:self-consistency} are coupled through $m$ and are analytic, each solution for $m^+$ comes with its own solution for $m^-$, so a fold bifurcation for either population always incurs a concomitant pair of solutions for the other, leading to multiple folds as are characteristic of the binomial RFIM. Here, we are only discussing the ``large'' fold bifurcations that are not squeezed against mobility of 0 or 1.} We show an example of such a dynamical outcome in Figure~\ref{fig:stability}f, where the ($-$) population is almost completely located in one out of two regions, and the ($+$) population likewise collects.

Additionally, a discontinuous transition can lead coupled regions to sustain intermittent dynamics as they are repeatedly ``driven'' across the fold bifurcation by stochastic migrant flow. 
An illustrative example is of region a sitting on the upper solution manifold while coupled to region b sitting on the lower mobility manifold. Region a will relax to lower mobility as migrants leave. At the same time, it will pump region b to the right on its solution branch by increasing $n_b^+$. If region a crosses the bifurcation to drop to the lower branch of the manifold simultaneously with the jump up of region b, then this cycle can repeat. The condition of balance in the number of ($+$) individuals assumed by this picture for two regions, such that the drop and the jump are simultaneous, arises from conservation of the ($+$) population across the two regions and of the ($-$) population within each region. If such a constraint is not imposed, stochastic fluctuations could still drive switching, but less regularly. This point emphasizes how such dynamics require the system to be poised in the right parts of parameter space and how changes in population composition, a feature not usually discussed in physics of the RFIM, manifests in the dynamics.

\begin{figure}
\centering
	\includegraphics[width=\linewidth]{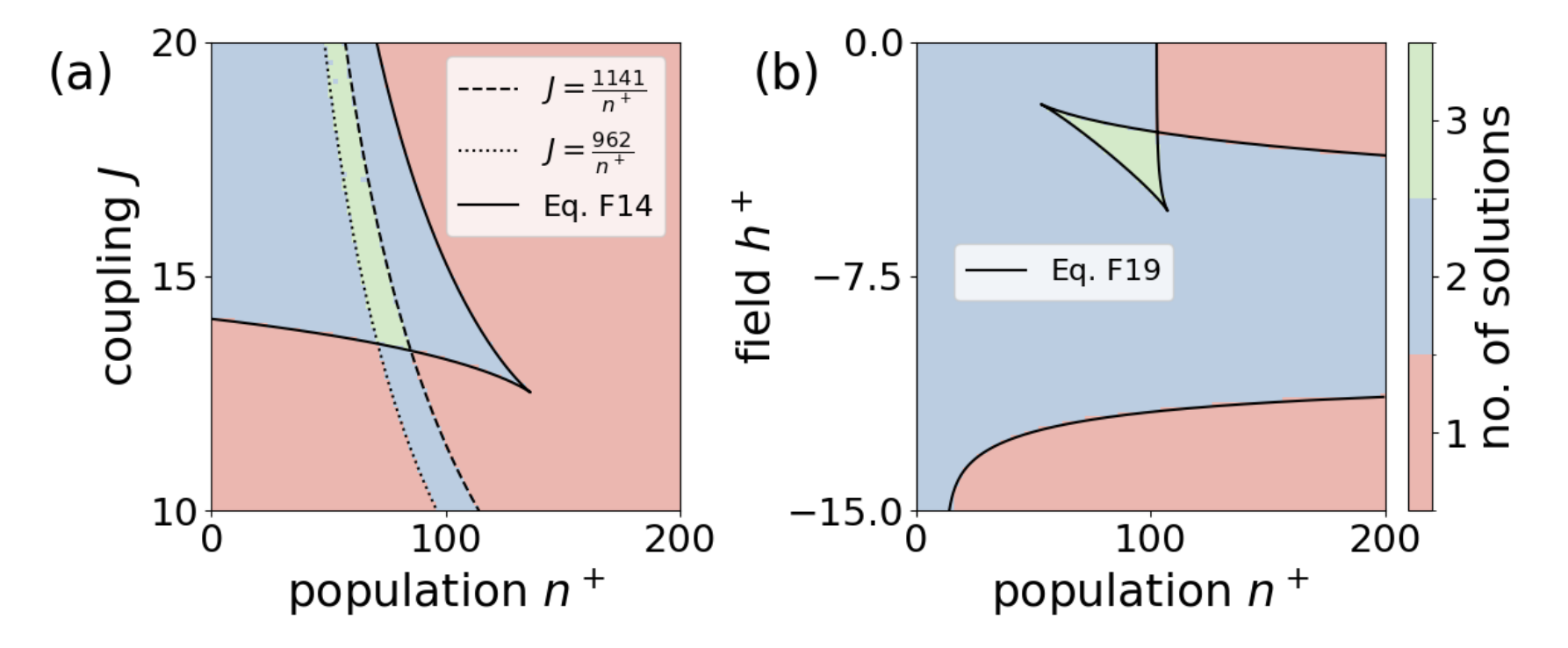}
\caption{Solution landscapes as a function of the (a) coupling strength $J$ and (b) field $h^+$ against the population size $n^+$, effectively a parameter that drives regions into new regimes. Color indicates number of stable solutions. The black solution boundaries are obtained using analytical fold conditions (Appendix \ref{sec:sol_boundary}). Parameters $n=200$, $h^+=-2.52$, $h^-=-10.52$ on the left, and parameters $n=200$, $h^-=-10.52$, $J=15$ on the right.}\label{fig:solution landscape}
\end{figure}

To find the locations of parameter space where we expect switching between branches of the solution manifold, we map the solution landscape using numerical calculations. In Figure~\ref{fig:solution landscape}, we highlight the fold bifurcations by enumerating the number of stable solutions with changing couplings $J$, fields $h^+$, and the number $n^+$.\footnote{A fold bifurcation requires that the ($+$) population satisfy $h^+ < -2$ and $Jn^+/n \geq 4$. This is shown in Appendix \ref{sec:min_hp}.} The colors indicate where only one solution can be found for both ($+$) and ($-$) populations (red), where they have two solutions (blue), and where the fold bifurcations for both populations overlap, leading to the unusual three-solution configuration (green).\footnote{This corresponds to an extra bend in the curve defined on the right hand side of Eq~\eqref{eq:self-consistency}. Intuitively, the distance between the two fields means that the sigmoidal curve does not show a single transition from convexity to concavity, but multiple from the two fields. This suggests a scheme for generalizing the intuition from the binomial to a multinomial model. Note that this will not happen for a unimodal distribution of disordered fields, e.g., Gaussian.} 
The solution boundaries are where folds emerge in the equation for the self-consistency condition and so moving along a boundary is equivalent to holding fixed the number of zero-points as we vary the parameters. We obtain a set of analytical fold conditions (black lines in Figure~\ref{fig:solution landscape}, described further in Appendix~\ref{sec:sol_boundary}) that capture how the topology of the self-consistency condition is conserved.

As we approach regions of bistability, we find transients that show intermittent fluctuations in migration rate. Consider first dependence on the ratio of $n^+$ to $n^-$. If we approach a fold bifurcation from the right side, where $n^+$ is larger than the value required to initialize the regions symmetrically between the ends of the fold bifurcation, random but rare fluctuations will drive one of the manifolds across the fold bifurcation to the bottom branch for a short time. These lead to intermittent drops in migration out from one of the regions (Figure~\ref{fig:intermittency high}). We can also approach the fold from below; here, intermittency drives brief increases in migration outflow from the region with high mobility and echoing the data examples in Figure~\ref{fig:data} (see Figure~\ref{fig:intermittency low} for simulations).\footnote{It is clear that the width of the fold bifurcation $w$ determines how long it takes for such spontaneous switching to the other solution branch, and so the most interesting regions are for physically reasonable times $t$ where the size of migrant fluctuations is sufficient to drive the system to instability, or very roughly $t \sim (w/2\sigma)^2$ for $\sigma \sim \sqrt{fn^+m^+}$. This time is much shorter for regions initialized with mobility rightwards of the center of the fold bifurcation because this favors larger $m^+$ and thus larger fluctuations in migrant number, whereas regions to the left preferentially fall into a transient immobile phase.} 
Similar boundaries to marginal stability could be approached with fields and couplings. Deep in the red regions, where there is only a single solution and the solution manifold is continuous, the regional populations will move towards a stable long-term configuration with small, constrained fluctuations away from steady state. 

When $n^-$ is variable, oscillations compete with another dynamical mode.  
Consider the fluctuation correlations between inflow $\delta(t) = \sum_{\rm b\neq a} n_{\rm b}(t) m_{\rm b}(t) / d_{\rm b}^{\rm out}$ and outflow $\delta'(t) = n_{\rm a} m_{\rm a}$ from region a. We calculate these when displaced by a time $\Delta t$, or the time-correlation function $\langle\delta (t) \delta '(t+\Delta t)\rangle$. As we show in Figure~\ref{fig:large graphs}, this function shows a negative peak at $\Delta t=0$ and is symmetric about that point as we would expect from a system that cycles between the upper and lower manifolds of the solution. The correlation, however, is weak because two regions in a symmetric configuration (sharing the same mobility and population) is also stable. In other words, the cyclic dynamics compete with the tendency of the two regions to relax to a symmetric configuration with low migrant flow, which can then dominate the simulation when it takes a long time to diffuse out from it. 
For $r>2$ regions, we can generalize the two dynamical modes, either regular pulses in ($+$) migration or collective symmetry in mobility as a function of network structure.

\section{Waves and cascades in more than two regions}
Beyond the dyadic example, the structure of a large network determines the kind of dynamical stability we observe. Here, we focus on network structures that facilitate propagating waves and cascades of global mobility when regions are poised along the fold bifurcation.

\begin{figure}
\centering
\includegraphics[width=\linewidth]{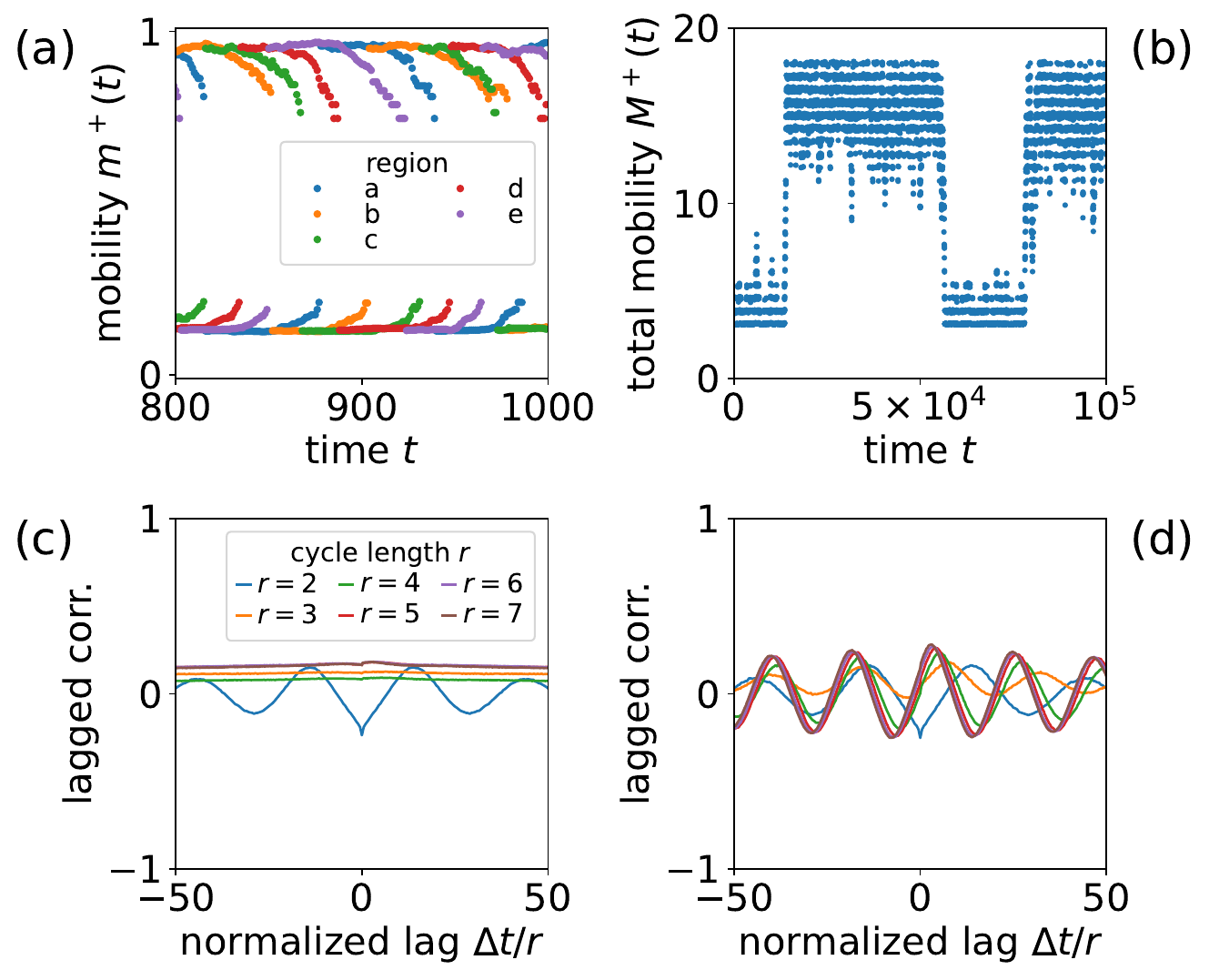}
\caption{(a) Traveling wave in a directed cycle of length $r=5$. (b) Intermittency in a fully connected graph of size $r=20$. (c) Emergence of propagating waves in cycles of $r$ regions indicated by time-lagged correlation coefficient between outflow vs.~inflow. Negative lag corresponds to case where outflow precedes inflow and positive lag where outflow follows inflow. Averaged over all regions for simulation time $t=5\times10^4$. (d) Time-lagged correlation coefficient for full graph shows no temporal symmetry breaking. For all panels, parameter values for fold bifurcation are $h^+=-2.52$, $h^-=-10.52$, $J=12$, $n^+=89$, $n=200$. }
\label{fig:large graphs}
\end{figure}

Directed cycles propagate wave fronts coherently, the simplest case of which is a single ring. As we show in Figure~\ref{fig:large graphs}, this difference is displayed in a characteristically different time correlation function for fluctuations in comparison with the $r=2$ case. Here, historical outflow is negatively correlated with current inflow (i.e., high inflow comes after a period of low outflow) and future outflow is positive correlated with current inflow (i.e., current inflow leaves coherently). This is exactly the picture of a traveling wave moving from one region to another along the entire cycle, and simulations show that such traveling ``shock waves'' emerge spontaneously from uniform initial conditions \cite{sugiyamaTrafficJams2008}, indicating how chains coherently funnel impulses between regions.

In contrast, large out-degrees would equally divide outflow amongst many neighbors, flattening large impulses. As we increase regional connectivity by adding routes, we might expect the suppression of large fluctuations as pulses are diluted. We find, however, the emergence of repeated cascades as the fully connected network switches between global configurations of high or low mobility. As we show in the flat inflow and outflow time-correlation function in Figure~\ref{fig:large graphs}, collective excitation and quiescence are a result of random, spontaneous cascades that drive the collective from one of the solution manifold to the other.

The collective states, where all regions occupy the same mobility solution manifold are stable, especially when the fold bifurcations are wide and the networks are large. 
We can see this by sketching an argument for the time it takes for a fully connected network to transit from a collectively low-mobility configuration to a high one, when $r\gg1$. Let us start from the stable, symmetric configuration on the lower mobility manifold. Taking all regions to be poised in the center of the fold bifurcation, migrant fluctuations from neighboring regions must overpower the restoring potential to drive a first region across the fold bifurcation.\footnote{To a first approximation, the time for the first region to cross the fold bifurcation is the combination of diffusive contributions from Poissonian in and outflows of migrants competing against the averaged linear restoring potential, or an Ornstein-Uhlenbeck process. A more refined approximation accounts for the increasing steepness of the mobility gradient at the saddle point. This means that the Ornstein-Uhlenbeck argument for crossing underestimates the expected time to cross to the high-mobility manifold. Numerical calculations show that when $r\gg1$, the expected time $\tau$ for the first region to cross the fold bifurcation approaches $\tau \sim w^2/8\sigma^2\log r$ (see Appendix~\ref{app:fpt_crossing}).}
The next region to switch to high mobility experiences both a positive migrant inflow from the single region of high mobility and along with diffusive kicks from the other $r-1$ regions. The competing (and presumably faster) timescale is for the single high mobility region to relax to the lower mobility manifold, a time that lengthen as additional regions join the high mobility branch. While few regions have switched to high mobility, the low mobility configuration is stable, but the competing timescales move in tandem as to increase the stability of the high-mobility regions and decrease the stability of the low-mobility regions as we show in Figure~\ref{fig:drift_transition}. If a critical point is crossed, where the timescale for switching up is faster, then the entire system moves quickly to high mobility. Then, the high-mobility configuration is stable, and the inverse argument applies for the collective transition to quiescence. We show in Figure~\ref{fig:large graphs} examples of these two dynamical modes with numerical simulations: total mobility $M^+(t) \equiv \sum_{{\rm a}=1}^r m_{\rm a}^+(t)$, summed over all $r$ regions, spends long durations at low or high values with relatively rapid transitions. Thus, we find a phenomenon of collective excitation and quiescence with times of rapid turnover as high or low mobility configurations cascade throughout the network.

\section{Shocks}
\begin{figure}
\centering
\includegraphics[width=\linewidth]{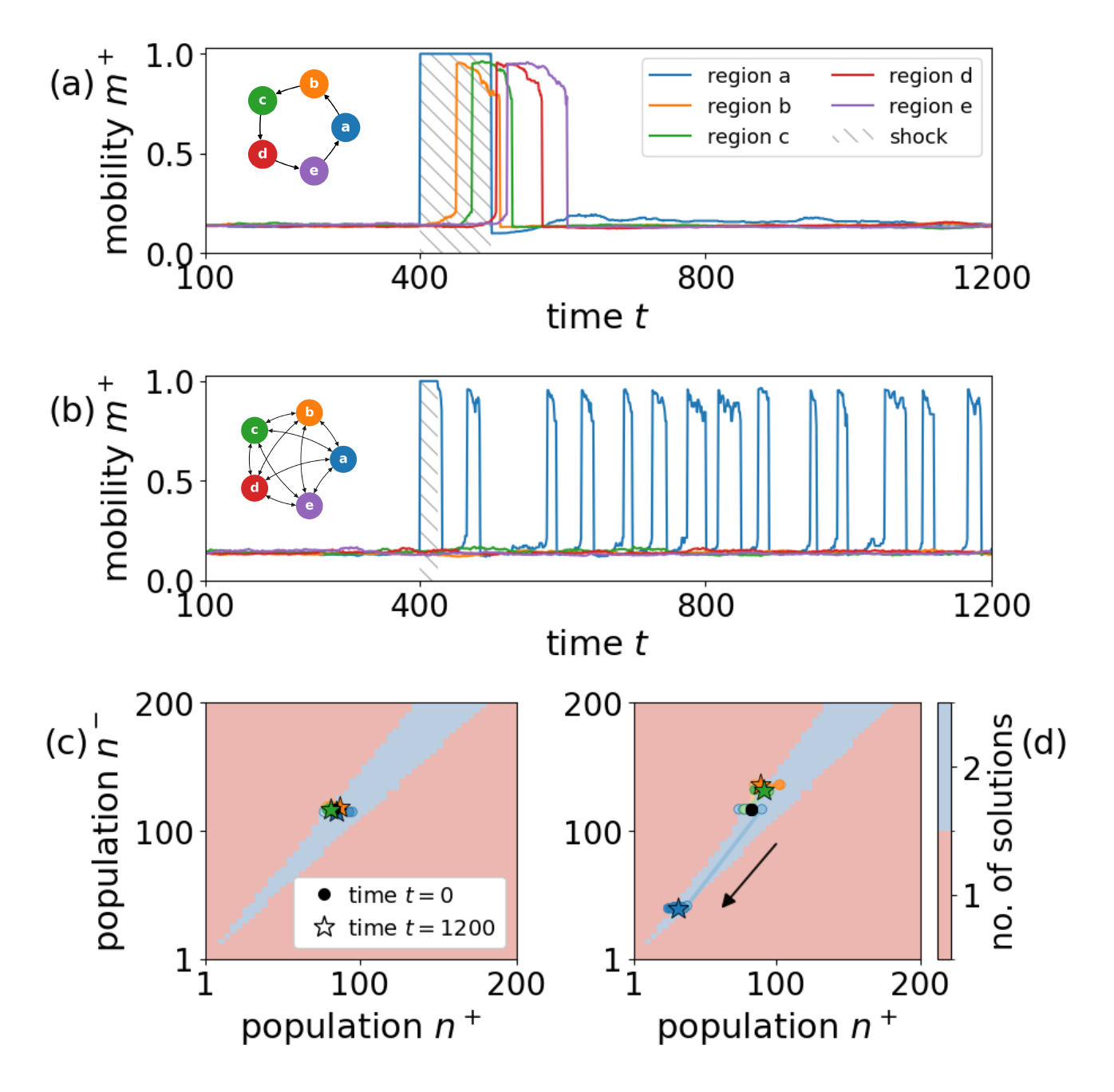}
\caption{Shock response in cyclic and fully connected networks with $r=5$ regions. Shocks applied to region~a (blue) at $t=400$. Shock duration (shaded region) is inversely proportional to the total outgoing migration rate, ensuring that approximately the same number of migrants are expelled in each case; thus, shocks are shorter in the fully connected topology than in the cyclic chain. Trajectories of regions~a, b, and~c are shown in blue, orange, and green respectively, in the $(n^+, n^-)$ plane; the black point marks the common initial position of every region and the colored star marks the final position of each region.
(a,\,c)~($+$) population shock, cyclic chain;
(b,\,d)~total population shock, fully connected.
The black arrow denotes the direction in which the shocked region travels in the $(n^+, n^-)$ plane.
Mobilities initially set at random. Parameters are $h^+=-2.52$, $h^-=-10.52$, $J=12$, $n=200$.}
\label{fig:shocks}
\end{figure}

We model a shock as an instantaneous mobilization of one region, holding for some shock duration $\tau$ a subset of migrants to $s_{\rm i}=1$ for a target region. Consider systems near a fold bifurcation with low ($-$) mobility $m^-\ll1$. Here, a temporary shock to the ($+$) subpopulation alone generally relaxes back to the equilibrium solution as we would expect from the dynamical stability argument above, an example of which we show in Figure~\ref{fig:shocks}a.\footnote{The relaxation is stronger in the fully connected case since the outgoing migrants are distributed among a larger number of regions as we show in the example of Figure~\ref{fig:shocks}.} A shock that redistributes both the ($+$) and ($-$) populations, however, produces qualitatively different dynamics as in Figure~\ref{fig:shocks}b. In contrast to the first shock, the latter translates the entire system diagonally down the $(n^+,n^-)$ plane as we show in Figures~\ref{fig:shocks}c and d. The thus translated system does not recover quickly because the ($-$) population is immobile, and this means that the shocked region sustains a memory of the shock. Since the fold bifurcation (blue region) thins with total population, the mobility is susceptible to smaller fluctuations, and we find the emergence of intermittent, large migrant flows out from the depopulated region. Regions away from a fold bifurcation, in comparison, do not show any qualitative change (Figure \ref{fig:shocks_SI}). Thus, even short shocks can leave a long-term trace in both the total population and its dynamical sensitivity.

\section{Noise in collective decisions}\label{sec:automaton}
Besides fluctuations in migrant flow, the process via which the migrants in a region equilibrate is in principle stochastic, an aspect excluded from the mean-field solution to the mobility in Eq~\eqref{eq:self-consistency} and thus in the dynamics that follow from Eq~\eqref{eq:dn stochastic}. To account for the additional source of noise, we construct a stochastic automaton simulation alongside the stochastic dynamics described in Eq~\eqref{eq:dn mft}, where migrants repeatedly interact with local neighbors to ``decide'' on their mobility; this takes into account ``thermal'' noise, a fundamentally different source of instability that reflects the shape of the energy landscape, as fluctuations in $n^+$ and $n^-$ move us from one energy landscape to another.

\begin{figure}
\centering
\includegraphics[width=\linewidth]{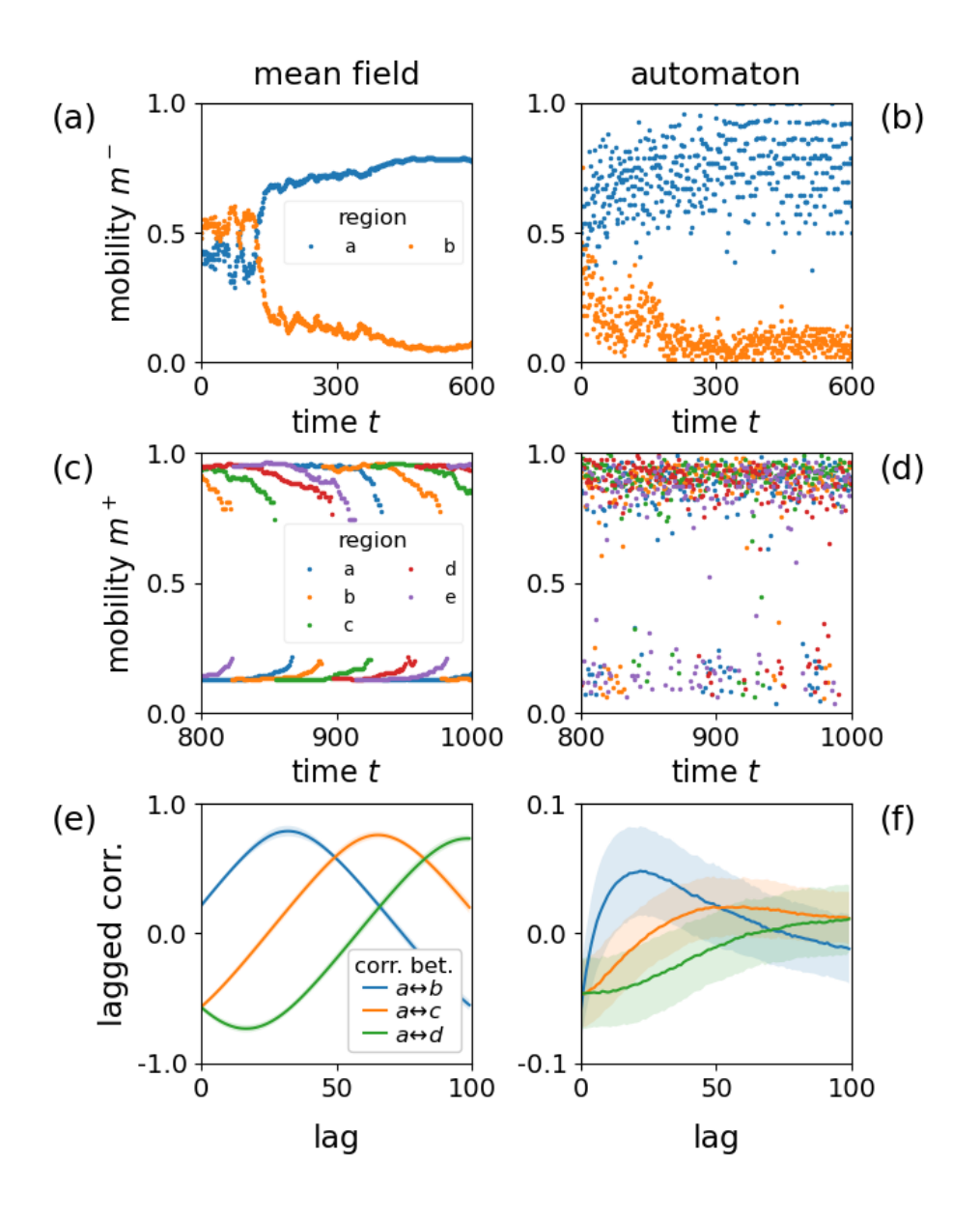}
\caption{Comparison of stochastic mean-field dynamics with automaton. (a, b) Spatial symmetry breaking in two regions and (c, d) traveling waves in five regions connected in cycle. (e, f) Normalized lagged correlation  $\langle \delta m_{\rm a}^+(t)\,\delta m_{\rm a'}^+(t+\Delta t)\rangle/\sigma_{m_{\rm a}^+}\sigma_{m_{\rm a'}^+}$ between regions $a$ and $a'$ in a cyclic chain configuration with $r=5$. Here shaded areas denote one standard deviation across 100 simulation runs. Parameters are $h^+=-2.52$, $h^-=-10.52$, $J=12$, $n=200$.}
\label{fig:automaton}
\end{figure}

Thermal fluctuations have two consequences. First, the two branches of the mean-field solution are occupied at differential rates given by the difference in free energies, which can lead to stark imbalance in the occupation rates of either branch in the thermodynamic limit. 
Then, the dynamical outcomes of bifurcations in the mean-field model solutions emerge most strongly when there is no single dominant global energy minimum. Second, thermal fluctuations diffuse initially coherent pulses of migrants. As a result, traveling waves do not persist as well over the cycle, even if the region-to-region fluctuation autocorrelation functions show signatures of pulse travel as in Figure~\ref{fig:automaton}b. Only when temperature is low, the mobility sharply defined about the minimum of the energy landscape, and the free energies of solution branches similar can we obtain the sharp fold bifurcation boundaries that allow for mobility pulses to travel coherently through a cycle. Spontaneous symmetry breaking, however, is more robust and we can obtain examples of this outcome in simulation as in Figure~\ref{fig:automaton}a.

\section{Discussion}
Migration is a fundamentally dynamical phenomenon in which individuals' decisions change over time and shape emergence of collective migration patterns \cite{oh2022interplay}. Yet, largely missing from the discussion of models is how migration patterns change over time as a result of interacting, individual-level decisions. We frame this question as one of nonlinear dynamics by building a minimal migration model (Figure~\ref{fig:model diagram}). In the minimal model, we use a binomial distribution of fields to indicate more- and less-mobile migrants. Then, its mean-field solution represents the way decisions are on average reached within each region, and the regions are themselves coupled in a network as migrants move from one to another. The approach thus combines collective decisions at both individual (through influence) and regional levels (through migration) to show how global migration patterns feed back to individual-level behavior.

While steady-state outcomes akin to the key assumptions of the push-and-pull framework characterize a large part of the parameter space, the model also displays symmetry breaking, spatially (meaning across regions as in Figure~\ref{fig:stability}) and temporally as in Figure~\ref{fig:large graphs}. Spatial symmetry breaking consists of differentiation between long-term regional populations despite symmetry in the equations and initial conditions. This phenomenon reflects the tendency of less-mobile populations to become more stable when they increase in number and less stable when they decrease in number, a direct consequence of social influence. This drives them to naturally ``evaporate'' from the less populated regions and ``condense'' in more populated regions\footnote{We show an example of symmetry breaking in migrant flows, as would result from symmetry breaking in populations, in Appendix \ref{sec:data_dynamics} for Somalia.}. This prediction from first principles presents a general mechanism (i.e., interaction and mixed populations) via which asymmetry in population sizes in settled areas emerges \cite{zipfHumanBehavior1949}.

Temporal symmetry breaking, on the other hand, in the form of recurrent waves of migration, is an outcome of both bistability and precise, symmetric placement of the migration regions within the fold bifurcation. Such placement (if it were to be a plausible explanation of periodic waves of migration) would require a mechanism that drives migration to these liminal spaces. Importantly, the fold bifurcation also is surrounded by a larger region of parameter space that displays intermittent fluctuations as regions cross the fold bifurcation because of a random, spontaneous jump in incoming migrants and then relax. When we look at displacement statistics, we see patterns similar to regions sitting to the left of the fold bifurcation, as if they were being driven up to short periods of high outmigration from incoming flows. This outcome provides one possible explanation of both recurrent cycles of migration and large fluctuations without resorting to complex institutional, economic, or exogenous drivers.

Besides these predictions, we also show that the structure of the migration flow networks determines the types of global migration dynamics when there is bistability. The restriction, for example, of migration routes into chain structures supports coherent propagation, thus increasing the size of the shock downstream. Such ``funnels'' can lead to traveling migration waves from one region to another. Complementarily, adding outgoing routes to multiple regions blunts waves by distributing any single shock below the threshold at which it can drive a neighbor across critical points. Adding more routes in general, however, does not simply reduce major shifts in migration. Surprisingly, it may not obviate spikes but can drive the entire network to collective excitation with mass migration all around. Since collective quiescence separates these periods of excitation, the high and low mobilities represent stable dynamical configurations of the whole system, and the switching between the two modes is sudden and unpredictable. These observations of collective modes focus less on the incentives of the individual, such as are emphasized in political debates, but show how structural considerations like the migration routes have substantial, emergent, and surprising impacts on global outcomes. 

There are a couple directions in which it would be fruitful to consider beyond the limiting assumptions of the proposed framework. First, disorder in the edges such that certain routes allow for more capacity than others would break the symmetry in the steady-state population distributions over the regions. It would create more variety in terms of the long-term dynamics, but it does not seem likely to introduce new types of dynamics. In contrast, we do not consider the attractive appeal of a region, which could lead to asymmetric flows between regions and nonequilibrium outcomes.  Another aspect is time correlations or non-Poissonian migration flows. Indeed, we find that individuals tend to group when moving, such as in Somalian displacement data, so this may affect the onset of dynamical transitions by exacerbating fluctuations. Finally, decisions are most likely to be influenced by a small group of trusted individuals and not by everyone in the region as we assume in the mean-field framework; in other words, social network structure often matters \cite{barabasiMeanfieldTheory1999, easleyNetworksCrowds2010, newmanNetworks2018}. These present variations on the mean-field framework that we develop here, which sets up a much richer set of theoretical extensions to be considered.

The framing of migration as a collective outcome also seems relevant for the recently popular topic of migration control. Often, this is framed in terms of individual choices to migrate and thus control about changing individual-level incentives, opportunities, and heuristics \cite{sjaastadCostsReturns1962, dehaasTheoryMigration2021, bijakBayesianModelBased2022}. Once we account for collective decisions, however, it is clear that an effective policy would change the stable dynamical outcomes, a system-level property that would alter not only the propensity of all individuals in a region to migrate but also the pan-regional incidence of migration. In the model, such questions bring attention to ``control'' parameters like the relative of individuals unwilling to move (the ratio $n^+/n^-$) or social influence such as might be changed through the mechanisms of online information sharing. The latter is especially important because it determines the shape of the mobility manifold. If regions lie on the continuous manifold, changes in behavior change smoothly with imposed policy. If close to the discontinuous manifold, however, a policy would be most effective when in effect before a crossover through the transition. If a region, for example, passes this tipping point, then migration will be difficult to restore to its previous value; of course, tipping could also be leveraged to instigate change. These qualitative insights suggest that it is critical to measure the migration response curve of the population, a yet unconsidered idea in the literature. 

The main implication of the model for migration policy is that changes in migration do not have to be indicative of exogenous shocks but may be an endogenous outcome of collective decisions. Indeed, we show how interactions can lead to sharp transitions in local mobility. As a result, regions poised around such regions of parameter space would see large fluctuations in migration and proximity to such bifurcations would incur fluctuations with higher frequency. This aspect could confound correlational tests for measuring the impact of natural disasters and conflicts on migration patterns \cite{thalheimerLargeWeather2023}. 
Bifurcations are not the only source of intermittent fluctuations. As we show, migrant flow coupling between regions can also be locally unstable as small, stochastic fluctuations push regions away from marginally stable configurations (Figure~\ref{fig:stability}c).
Thus, it is the interplay between stochasticity and the existence of multiple solutions for mobility that are essential for the intermittent dynamics we find. The consequences of exogenous perturbations could only be disentangled from those of collective decisions once we could formulate their variety and distributions.


\begin{acknowledgments}
We thank Shlomo Havlin, Jan Korbel, Stefan Thurner, and Rudi Hanel for discussion. EDL acknowledges funding in part by the Austrian Science Fund (FWF) 10.55776/ESP127 and by the Austrian Federal Ministry for Innovation, Mobility and Infrastructure (BMIMI) as part of the project GZ 2021-0.664.668. WSO is pleased to acknowledge the support from the Natural Sciences and Engineering Research Council of Canada Discovery Grant under award \# RGPIN-2026-05104. NK acknowledges Österreichische Forschungsgemeinschaft (ÖFG) for their travel grant (IK-00001774). 
\end{acknowledgments}

\bibliographystyle{unsrt}
\bibliography{biblio,refs} 

@article{andersonGravityModel2011,
  title = {The {{Gravity Model}}},
  author = {Anderson, James E.},
  year = 2011,
  month = sep,
  journal = {Annual Review of Economics},
  volume = {3},
  number = {1},
  pages = {133--160},
  issn = {1941-1383, 1941-1391},
  doi = {10.1146/annurev-economics-111809-125114},
  urldate = {2026-03-24},
  langid = {english}
}

@article{barabasiMeanfieldTheory1999,
  title = {Mean-Field Theory for Scale-Free Random Networks},
  author = {Barabasi, Albert-Laszlo and Albert, Reka and Jeong, Hawoong},
  year = 1999,
  journal = {Physica A},
  volume = {272},
  pages = {173--187},
  publisher = {Elsevier},
  doi = {10.1016/S0378-4371(99)00291-5},
  langid = {english}
}

@article{beyerGravityModels2022,
  title = {Gravity Models Do Not Explain, and Cannot Predict, International Migration Dynamics},
  author = {Beyer, Robert M. and Schewe, Jacob and {Lotze-Campen}, Hermann},
  year = 2022,
  month = feb,
  journal = {Humanities and Social Sciences Communications},
  volume = {9},
  number = {1},
  pages = {56},
  issn = {2662-9992},
  doi = {10.1057/s41599-022-01067-x},
  urldate = {2026-05-15},
  langid = {english}
}

@article{bialekSocialInteractions2014,
  title = {Social Interactions Dominate Speed Control in Poising Natural Flocks near Criticality},
  author = {Bialek, W. and Cavagna, A. and Giardina, I. and Mora, T. and Pohl, O. and Silvestri, E. and Viale, M. and Walczak, A. M.},
  year = 2014,
  month = may,
  journal = {Proceedings of the National Academy of Sciences},
  volume = {111},
  number = {20},
  pages = {7212--7217},
  issn = {0027-8424, 1091-6490},
  doi = {10.1073/pnas.1324045111},
  urldate = {2019-05-02},
  langid = {english}
}

@techreport{bijakAssessingUncertain2020,
  title = {Assessing {{Uncertain Migration Futures}}: {{A Typology}} of the {{Unknown}}},
  author = {Bijak, Jakub and Czaika, Mathias},
  year = 2020,
  number = {Deliverable 1.1},
  institution = {{University of Southampton and Danube University Krems}},
  langid = {english}
}

@book{bijakBayesianModelBased2022,
  title = {Towards {{Bayesian Model-Based Demography}}: {{Agency}}, {{Complexity}} and {{Uncertainty}} in {{Migration Studies}}},
  shorttitle = {Towards {{Bayesian Model-Based Demography}}},
  author = {Bijak, Jakub},
  year = 2022,
  series = {Methodos {{Series}}},
  volume = {17},
  publisher = {Springer International Publishing},
  address = {Cham},
  doi = {10.1007/978-3-030-83039-7},
  urldate = {2025-02-11},
  copyright = {https://creativecommons.org/licenses/by/4.0},
  isbn = {978-3-030-83038-0 978-3-030-83039-7},
  langid = {english}
}

@article{bircanEditorialNew2025,
  title = {Editorial: {{New}} Methodological Approaches for Migration and Mobility Studies: From Traditional to Big Data},
  shorttitle = {Editorial},
  author = {Bircan, Tuba and Qi, Haodong},
  year = 2025,
  month = oct,
  journal = {Frontiers in Human Dynamics},
  volume = {7},
  pages = {1710558},
  issn = {2673-2726},
  doi = {10.3389/fhumd.2025.1710558},
  urldate = {2026-05-15},
  langid = {english}
}

@article{cappelenForecastingImmigration2015,
  title = {Forecasting {{Immigration}} in {{Official Population Projections Using}} an {{Econometric Model}}},
  author = {Cappelen, {\AA}dne and Skjerpen, Terje and T{\o}nnessen, Marianne},
  year = 2015,
  month = dec,
  journal = {International Migration Review},
  volume = {49},
  number = {4},
  pages = {945--980},
  issn = {0197-9183, 1747-7379},
  doi = {10.1111/imre.12092},
  urldate = {2026-05-15},
  langid = {english}
}

@article{carlingMigrationAge2002,
  title = {Migration in the Age of Involuntary Immobility: {{Theoretical}} Reflections and {{Cape Verdean}} Experiences},
  shorttitle = {Migration in the Age of Involuntary Immobility},
  author = {Carling, J{$\oslash$}rgen},
  year = 2002,
  month = jan,
  journal = {Journal of Ethnic and Migration Studies},
  volume = {28},
  number = {1},
  pages = {5--42},
  issn = {1369-183X, 1469-9451},
  doi = {10.1080/13691830120103912},
  urldate = {2026-03-24},
  langid = {english}
}

@article{castellanoStatisticalPhysics2009,
  title = {Statistical Physics of Social Dynamics},
  author = {Castellano, Claudio and Fortunato, Santo and Loreto, Vittorio},
  year = 2009,
  month = may,
  journal = {Reviews of Modern Physics},
  volume = {81},
  number = {2},
  pages = {591--646},
  issn = {0034-6861, 1539-0756},
  doi = {10.1103/RevModPhys.81.591},
  urldate = {2019-05-15},
  langid = {english}
}

@article{cohenInternationalMigration2008,
  title = {International Migration beyond Gravity: {{A}} Statistical Model for Use in Population Projections},
  shorttitle = {International Migration beyond Gravity},
  author = {Cohen, Joel E. and Roig, Marta and Reuman, Daniel C. and GoGwilt, Cai},
  year = 2008,
  month = oct,
  journal = {Proceedings of the National Academy of Sciences},
  volume = {105},
  number = {40},
  pages = {15269--15274},
  issn = {0027-8424, 1091-6490},
  doi = {10.1073/pnas.0808185105},
  urldate = {2026-05-15},
  langid = {english}
}

@article{couzinEffectiveLeadership2005,
  title = {Effective Leadership and Decision-Making in Animal Groups on the Move},
  author = {Couzin, Iain D. and Krause, Jens and Franks, Nigel R. and Levin, Simon A.},
  year = 2005,
  month = feb,
  journal = {Nature},
  volume = {433},
  number = {7025},
  pages = {513--516},
  issn = {0028-0836, 1476-4687},
  doi = {10.1038/nature03236},
  urldate = {2020-04-01},
  langid = {english}
}

@article{couzinUninformedIndividuals2011,
  title = {Uninformed {{Individuals Promote Democratic Consensus}} in {{Animal Groups}}},
  author = {Couzin, Iain D and Ioannou, Christos C and Demirel, G{\"u}ven and Gross, Thilo and Torney, Colin J and Hartnett, Andrew and Conradt, Larissa and Levin, Simon A and Leonard, Naomi E},
  year = 2011,
  month = dec,
  journal = {Science},
  volume = {334},
  langid = {english}
}

@article{czaikaMigrationDecisionMaking2021,
  title = {Migration {{Decision-Making}} and {{Its Key Dimensions}}},
  author = {Czaika, Mathias and Bijak, Jakub and Prike, Toby},
  year = 2021,
  month = sep,
  journal = {The Annals of the American Academy of Political and Social Science},
  volume = {697},
  number = {1},
  pages = {15--31},
  issn = {0002-7162, 1552-3349},
  doi = {10.1177/00027162211052233},
  urldate = {2025-02-11},
  langid = {english}
}

@incollection{czaikaMigrationDrivers2022,
  title = {Migration {{Drivers}}: {{Why Do People Migrate}}?},
  shorttitle = {Migration {{Drivers}}},
  booktitle = {Introduction to {{Migration Studies}}},
  author = {Czaika, Mathias and Reinprecht, Constantin},
  editor = {Scholten, Peter},
  year = 2022,
  pages = {49--82},
  publisher = {Springer International Publishing},
  address = {Cham},
  doi = {10.1007/978-3-030-92377-8_3},
  urldate = {2026-03-24},
  isbn = {978-3-030-92376-1 978-3-030-92377-8},
  langid = {english}
}

@article{dehaasTheoryMigration2021,
  title = {A Theory of Migration: The Aspirations-Capabilities Framework},
  shorttitle = {A Theory of Migration},
  author = {{de Haas}, Hein},
  year = 2021,
  month = feb,
  journal = {Comparative Migration Studies},
  volume = {9},
  number = {1},
  pages = {8},
  issn = {2214-594X},
  doi = {10.1186/s40878-020-00210-4},
  urldate = {2024-06-11}
}

@book{easleyNetworksCrowds2010,
  title = {Networks, {{Crowds}}, and {{Markets}}: {{Reasoning About}} a {{Highly Connected World}}},
  author = {Easley, David and Kleinberg, Jon},
  year = 2010,
  publisher = {Cambridge University Press},
  isbn = {978-1-139-49030-6},
  langid = {english}
}

@article{epsteinInformationalCascades2002,
  title = {Informational {{Cascades}} and {{Decision}} to {{Migrate}}},
  author = {Epstein, Gil S.},
  year = 2002,
  journal = {SSRN Electronic Journal},
  issn = {1556-5068},
  doi = {10.2139/ssrn.304442},
  urldate = {2026-03-24},
  langid = {english}
}

@misc{gaskinDeepLearning2025,
  title = {Deep Learning Four Decades of Human Migration},
  author = {Gaskin, Thomas and Abel, Guy J.},
  year = 2025,
  month = jul,
  number = {arXiv:2506.22821},
  eprint = {2506.22821},
  primaryclass = {cs},
  publisher = {arXiv},
  doi = {10.48550/arXiv.2506.22821},
  urldate = {2026-02-26},
  archiveprefix = {arXiv},
  langid = {english}
}

@article{ghorbaniFlee32024,
  title = {Flee 3: {{Flexible}} Agent-Based Simulation for Forced Migration},
  shorttitle = {Flee 3},
  author = {Ghorbani, Maziar and Suleimenova, Diana and Jahani, Alireza and Saha, Arindam and Xue, Yani and Mintram, Kate and Anagnostou, Anastasia and Tas, Auke and Low, William and Taylor, Simon J.E. and Groen, Derek},
  year = 2024,
  month = sep,
  journal = {Journal of Computational Science},
  volume = {81},
  pages = {102371},
  issn = {18777503},
  doi = {10.1016/j.jocs.2024.102371},
  urldate = {2025-12-03},
  langid = {english}
}

@article{greiderNeighboringPatterns1985,
  title = {Neighboring {{Patterns}}, {{Social Support}}, and {{Rapid Growth}}: {{A Comparison Analysis}} from {{Three Western Communities}}},
  author = {Greider, Thomas and Krannich, Richard S.},
  year = 1985,
  journal = {Sociological Perspectives},
  volume = {28},
  number = {1},
  pages = {51--70},
  publisher = {Pacific Sociological Assn.},
  langid = {english}
}

@article{hoffmannMetaanalysisCountrylevel2020,
  title = {A Meta-Analysis of Country-Level Studies on Environmental Change and Migration},
  author = {Hoffmann, Roman and Dimitrova, Anna and Muttarak, Raya and Crespo Cuaresma, Jesus and Peisker, Jonas},
  year = 2020,
  month = oct,
  journal = {Nature Climate Change},
  volume = {10},
  number = {10},
  pages = {904--912},
  issn = {1758-678X, 1758-6798},
  doi = {10.1038/s41558-020-0898-6},
  urldate = {2026-01-05},
  langid = {english}
}

@article{jaynesInformationTheory1957,
  title = {Information {{Theory}} and {{Statistical Mechanics}}},
  author = {Jaynes, E. T.},
  year = 1957,
  month = may,
  journal = {Physical Review},
  volume = {106},
  number = {4},
  pages = {620--630},
  issn = {0031-899X},
  doi = {10.1103/PhysRev.106.620},
  urldate = {2019-05-03},
  langid = {english}
}

@article{leeStatisticalMechanics2015,
  title = {Statistical {{Mechanics}} of the {{US Supreme Court}}},
  author = {Lee, Edward D. and Broedersz, Chase P. and Bialek, William},
  year = 2015,
  month = jul,
  journal = {Journal of Statistical Physics},
  volume = {160},
  number = {2},
  pages = {275--301},
  issn = {0022-4715, 1572-9613},
  doi = {10.1007/s10955-015-1253-6},
  urldate = {2019-05-02},
  copyright = {All rights reserved},
  langid = {english}
}

@article{macdonaldChainMigration1964,
  title = {Chain {{Migration Ethnic Neighborhood Formation}} and {{Social Networks}}},
  author = {MacDonald, John S. and MacDonald, Leatrice D.},
  year = 1964,
  month = jan,
  journal = {The Milbank Memorial Fund Quarterly},
  volume = {42},
  number = {1},
  eprint = {3348581},
  eprinttype = {jstor},
  pages = {82},
  issn = {00263745},
  doi = {10.2307/3348581},
  urldate = {2024-05-24},
  langid = {english}
}

@book{newmanNetworks2018,
  title = {Networks},
  author = {Newman, Mark},
  year = 2018,
  publisher = {OUP Oxford},
  isbn = {978-0-19-252749-3}
}

@article{parkGeneralizedGravity2018,
  title = {Generalized Gravity Model for Human Migration},
  author = {Park, Hye Jin and Jo, Woo Seong and Lee, Sang Hoon and Kim, Beom Jun},
  year = 2018,
  month = sep,
  journal = {New Journal of Physics},
  volume = {20},
  number = {9},
  pages = {093018},
  issn = {1367-2630},
  doi = {10.1088/1367-2630/aade6b},
  urldate = {2024-06-16},
  langid = {english}
}

@article{qiModellingPredicting2023,
  title = {Modelling and Predicting Forced Migration},
  author = {Qi, Haodong and Bircan, Tuba},
  editor = {De Benedictis, Luca},
  year = 2023,
  month = apr,
  journal = {PLOS ONE},
  volume = {18},
  number = {4},
  pages = {e0284416},
  issn = {1932-6203},
  doi = {10.1371/journal.pone.0284416},
  urldate = {2026-05-15},
  langid = {english}
}

@article{reichlovaCanMotivation2007,
  title = {Can {{Motivation Theory Explain Migration Decisions}}?},
  author = {Reichlov{\'a}, Nat{\'a}lie},
  year = 2007,
  month = jan,
  journal = {Prague Economic Papers},
  volume = {16},
  number = {1},
  pages = {70--85},
  issn = {12100455, 2336730X},
  doi = {10.18267/j.pep.298},
  urldate = {2025-02-12},
  copyright = {https://creativecommons.org/licenses/by-nc-nd/4.0/},
  langid = {english}
}

@article{schewelUnderstandingImmobility2020,
  title = {Understanding {{Immobility}}: {{Moving Beyond}} the {{Mobility Bias}} in {{Migration Studies}}},
  shorttitle = {Understanding {{Immobility}}},
  author = {Schewel, Kerilyn},
  year = 2020,
  month = jun,
  journal = {International Migration Review},
  volume = {54},
  number = {2},
  pages = {328--355},
  issn = {0197-9183, 1747-7379},
  doi = {10.1177/0197918319831952},
  urldate = {2026-02-26},
  langid = {english}
}

@book{schragNotFit2010,
  title = {Not {{Fit}} for {{Our Society}}: {{Immigration}} and {{Nativism}} in {{America}}},
  author = {Schrag, Peter},
  year = 2010,
  publisher = {University of California Press},
  address = {London}
}

@article{shanahanEffectsImmigrant1999,
  title = {The {{Effects}} of {{Immigrant Diversity}} and {{Ethnic Competition}} on {{Collective Conflict}} in {{Urban America}}: {{An Assessment}} of {{Two Moments}} of {{Mass Migration}}, 1869-1924 and 1965-1993},
  author = {Shanahan, Suzanne and Olzak, Susan},
  year = 1999,
  journal = {Journal of American Ethnic History},
  volume = {18},
  number = {3},
  eprint = {27502449},
  eprinttype = {jstor},
  pages = {40--64},
  langid = {english}
}

@article{siminiUniversalModel2012,
  title = {A Universal Model for Mobility and Migration Patterns},
  author = {Simini, Filippo and Gonz{\'a}lez, Marta C. and Maritan, Amos and Barab{\'a}si, Albert-L{\'a}szl{\'o}},
  year = 2012,
  month = apr,
  journal = {Nature},
  volume = {484},
  number = {7392},
  pages = {96--100},
  issn = {0028-0836, 1476-4687},
  doi = {10.1038/nature10856},
  urldate = {2024-06-28},
  copyright = {http://www.springer.com/tdm},
  langid = {english}
}

@article{sjaastadCostsReturns1962,
  title = {The {{Costs}} and {{Returns}} of {{Human Migration}}},
  author = {Sjaastad, Larry A.},
  year = 1962,
  month = oct,
  journal = {Journal of Political Economy},
  volume = {70},
  number = {5, Part 2},
  pages = {80--93},
  issn = {0022-3808, 1537-534X},
  doi = {10.1086/258726},
  urldate = {2026-05-15},
  langid = {english}
}

@article{stickImmigrantsSocial2024,
  title = {Immigrants' {{Social Relations}} with {{Neighbours}}: {{Does}} the {{Population Density}} of the {{Neighbourhood Matter}}?},
  shorttitle = {Immigrants' {{Social Relations}} with {{Neighbours}}},
  author = {Stick, Max and Schimmele, Christoph and Karpinski, Maciej and Arsenault, Am{\'e}lie},
  year = 2024,
  month = jun,
  journal = {Journal of International Migration and Integration},
  volume = {25},
  number = {2},
  pages = {861--885},
  issn = {1488-3473, 1874-6365},
  doi = {10.1007/s12134-023-01107-8},
  urldate = {2026-03-26},
  langid = {english}
}

@article{sugiyamaTrafficJams2008,
  title = {Traffic Jams without Bottlenecks---Experimental Evidence for the Physical Mechanism of the Formation of a Jam},
  author = {Sugiyama, Yuki and Fukui, Minoru and Kikuchi, Macoto and Hasebe, Katsuya and Nakayama, Akihiro and Nishinari, Katsuhiro and Tadaki, Shin-ichi and Yukawa, Satoshi},
  year = 2008,
  month = mar,
  journal = {New Journal of Physics},
  volume = {10},
  number = {3},
  pages = {033001},
  issn = {1367-2630},
  doi = {10.1088/1367-2630/10/3/033001},
  urldate = {2026-05-06},
  langid = {english}
}

@article{suleimenovaGeneralizedSimulation2017,
  title = {A Generalized Simulation Development Approach for Predicting Refugee Destinations},
  author = {Suleimenova, Diana and Bell, David and Groen, Derek},
  year = 2017,
  month = oct,
  journal = {Scientific Reports},
  volume = {7},
  number = {1},
  pages = {13377},
  issn = {2045-2322},
  doi = {10.1038/s41598-017-13828-9},
  urldate = {2025-12-03},
  langid = {english}
}

@misc{susmannBayesianProjection2025,
  title = {Bayesian {{Projection}} of {{Extant Refugee}} and {{Asylum Seeker Populations}}},
  author = {Susmann, Herbert and Raftery, Adrian E.},
  year = 2025,
  month = oct,
  number = {arXiv:2405.06857},
  eprint = {2405.06857},
  primaryclass = {stat},
  publisher = {arXiv},
  doi = {10.48550/arXiv.2405.06857},
  urldate = {2025-11-21},
  archiveprefix = {arXiv},
  langid = {english}
}

@article{thalheimerLargeWeather2023,
  title = {Large Weather and Conflict Effects on Internal Displacement in {{Somalia}} with Little Evidence of Feedback onto Conflict},
  author = {Thalheimer, Lisa and Schwarz, Moritz P. and Pretis, Felix},
  year = 2023,
  journal = {Global Environmental Change},
  volume = {79},
  pages = {102641},
  urldate = {2023-10-14}
}

@misc{unhcrPRMNDashboard2022,
  title = {{{PRMN Dashboard}}},
  author = {UNHCR},
  year = 2022,
  urldate = {2026-05-19}
}

@misc{unhcrSituationEurope2026,
  title = {Situation {{Europe Sea Arrivals}}},
  author = {UNHCR},
  year = 2026,
  urldate = {2025-11-20}
}

@misc{zensDynamicCount2025,
  title = {Dynamic {{Count Models}} with {{Flexible Innovation Processes}} for {{Irregular Maritime Migration}}},
  author = {Zens, Gregor and Bijak, Jakub},
  year = 2025,
  month = aug,
  number = {arXiv:2508.18716},
  eprint = {2508.18716},
  primaryclass = {stat},
  publisher = {arXiv},
  doi = {10.48550/arXiv.2508.18716},
  urldate = {2025-11-20},
  archiveprefix = {arXiv},
  langid = {english}
}

@book{zipfHumanBehavior1949,
  title = {Human {{Behavior}} and the {{Principle}} of {{Least Effort}}},
  author = {Zipf, George Kingsley},
  year = 1949,
  publisher = {Addison-Wesley},
  address = {Cambridge, MA}
}

@article{lee1966theory,
  title={A theory of migration},
  author={Lee, Everett S},
  journal={Demography},
  volume={3},
  number={1},
  pages={47--57},
  year={1966},
  publisher={Springer}
}

@article{plane1993demographic,
  title={Demographic influences on migration},
  author={Plane, David A},
  journal={Regional studies},
  volume={27},
  number={4},
  pages={375--383},
  year={1993},
  publisher={Taylor \& Francis}
}

@article{foster2017decomposing,
  title={Decomposing American immobility: Compositional and rate components of interstate, intrastate, and intracounty migration and mobility decline},
  author={Foster, Thomas B},
  journal={Demographic Research},
  volume={37},
  pages={1515--1548},
  year={2017},
  publisher={JSTOR}
}

@article{oh2022interplay,
  title={On the interplay among multiple factors: Effects of factor configuration in a proof-of-Concept migration agent-Based model},
  author={Oh, Woi Sok and Carmona-Cabrero, Alvaro and Mu{\~n}oz-Carpena, Rafael and Muneepeerakul, Rachata},
  journal={Journal of Artificial Societies and Social Simulation},
  volume={25},
  number={2},
  year={2022},
  publisher={JASSS}
}

@article{blair2026dynamics,
  title={Dynamics of internal displacement and conflict in Somalia},
  author={Blair, Christopher W},
  journal={Journal of Ethnic and Migration Studies},
  pages={1--35},
  year={2026},
  publisher={Taylor \& Francis}
}

@article{adam2025community,
  title={Community-based mortality surveillance among internally displaced vulnerable populations in Banadir region, Somalia, 2022--2023},
  author={Adam, Mohamed Hussein and Garba, Bashiru and Dahie, Hassan Abdullahi and Baruch, Joaquin and Polonsky, Jonathan A and Hassan, Jihaan and Mohamoud, Jamal Hassan and Ali, Dahir Abdi and Malik, SK Md Mamunur Rahman and Checchi, Francesco and others},
  journal={Frontiers in public health},
  volume={13},
  pages={1582558},
  year={2025},
  publisher={Frontiers Media SA}
}

@article{oh2024emergent,
  title={Emergent network patterns of internal displacement in Somalia driven by natural disasters and conflicts},
  author={Oh, Woi and Muneepeerakul, Rachata and Rubenstein, Daniel and Levin, Simon},
  journal={Global Environmental Change},
  volume={84},
  pages={102793},
  year={2024},
  publisher={Elsevier}
}

\clearpage
\appendix
\renewcommand{\thefigure}{S\arabic{figure}}
\renewcommand{\figurename}{Figure}
\setcounter{figure}{0}

\begin{figure}[b]\centering
	\includegraphics[width=\linewidth]{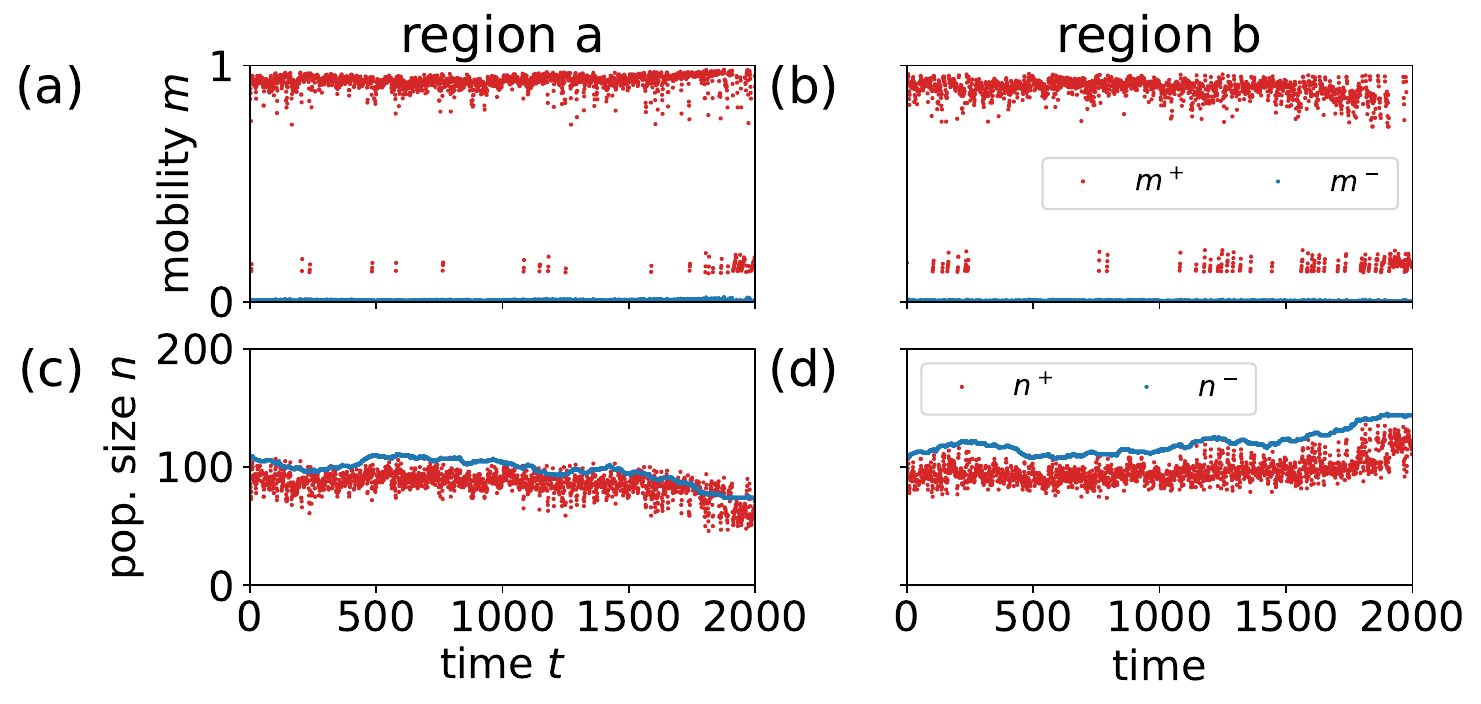}
	\caption{Intermittent fluctuations away from high-mobility state for regions positioned off-center from the fold bifurcation. We show the mean mobility of the populations $m^+(t)$ and $m^-(t)$ and respective populations $n^+$ and $n^-$.}\label{fig:intermittency high}
\end{figure}

\begin{figure}[b]\centering
	\includegraphics[width=\linewidth]{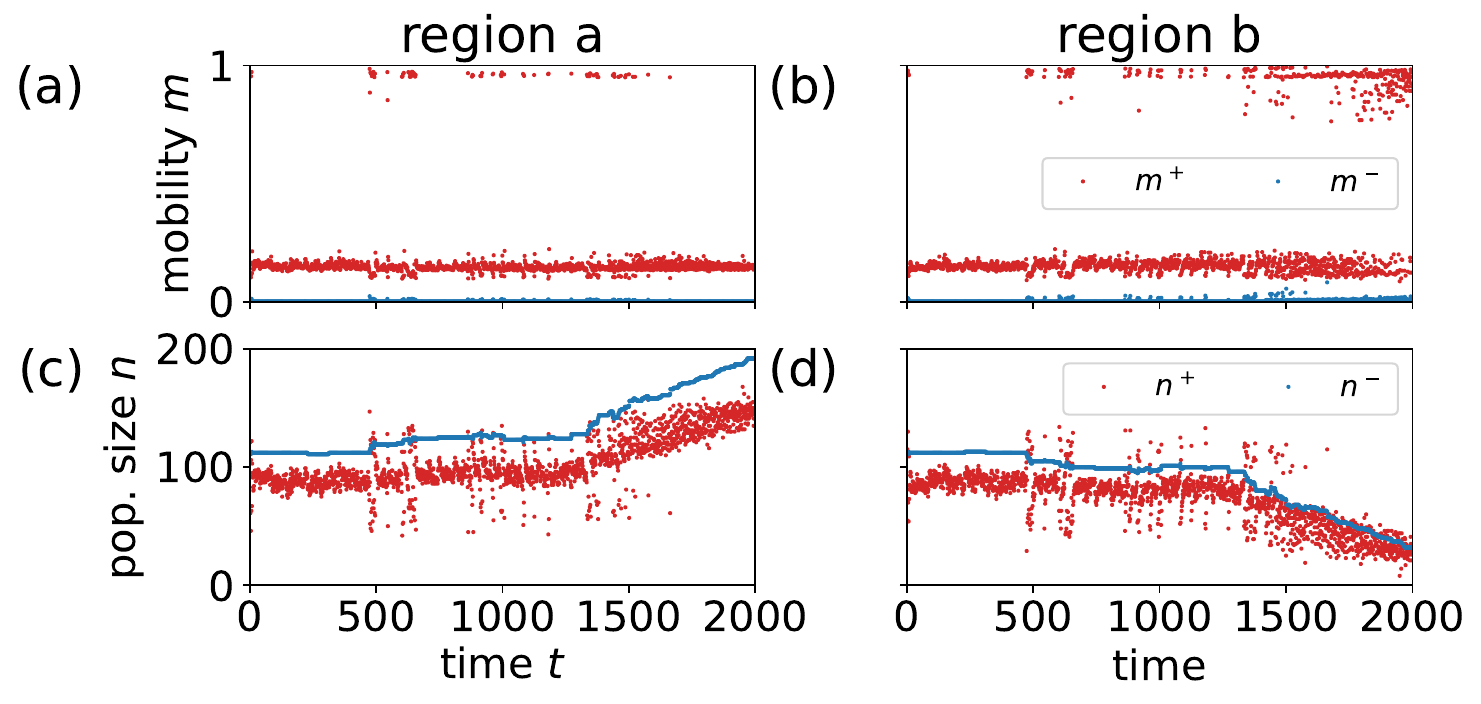}
	\caption{Intermittent fluctuations away from low-mobility state for regions positioned off-center from the fold bifurcation. We show the mean mobility of the populations $m^+(t)$ and $m^-(t)$ and respective populations $n^+$ and $n^-$.}\label{fig:intermittency low}
\end{figure}

\begin{figure}
\centering
\includegraphics[width=\linewidth]{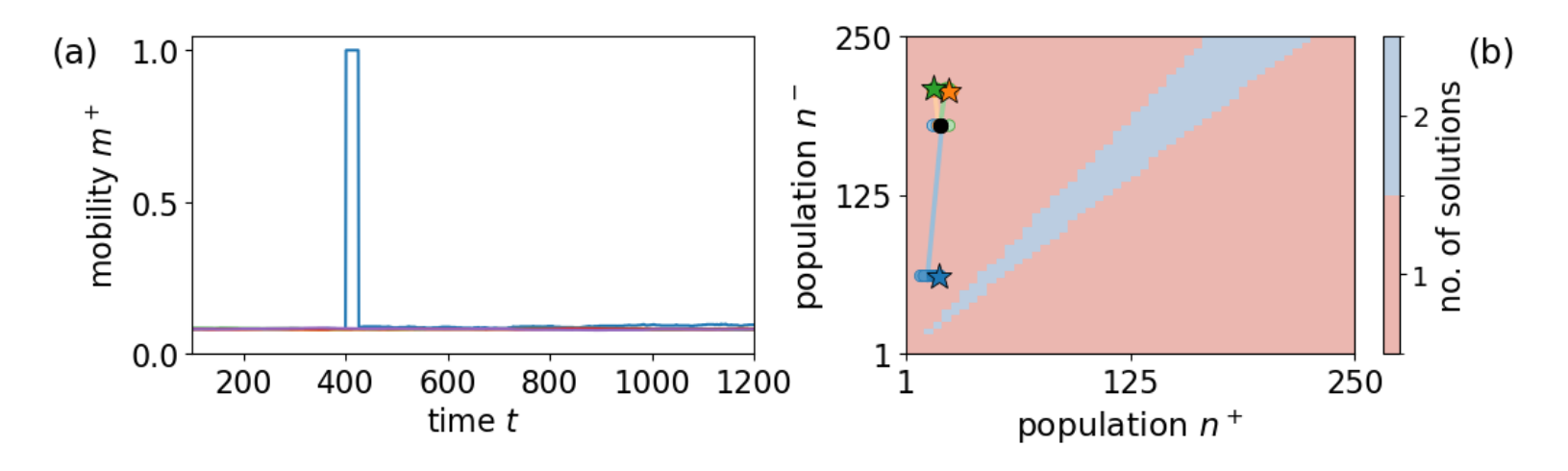}
\caption{Lack of interesting response to total population shock in fully connected topology with $r=5$ regions. Shock is applied to region~a (blue) from $t=400$ to $t=425$. Trajectories of regions~a, b, and~c are shown in blue, orange, and green respectively, in the $(n^+, n^-)$; black point marks the common initial position of every region and star marks the final.}
\label{fig:shocks_SI}
\end{figure}

\section{Note on automaton simulation}
We simulate the migration dynamics as a collection of coupled Ising models, with stochastic transport of spins between reservoirs superimposed on local Glauber-type relaxation. Each spin carries a fixed field label, corresponding to either the $h^-$ or $h^+$ subpopulation, and belongs at any given time to a single reservoir. Local interactions within each reservoir are mean-field, while migration between reservoirs is governed by a matrix of Poissonian transfer rates $f_{\rm ij}$. Here, we only consider uniform rates $f_{\rm ij}=f$.

The dynamics are implemented in discrete time with a sufficiently small increment $dt$, chosen so that at most one microscopic event is effectively resolved per update. Since there are $N$ spins in the system and only one spin is sampled uniformly at random at each elementary step, the microscopic event probabilities must be rescaled by the total number of spins. Accordingly, the equilibration rate $1/\tau$ and migration rates $f_{\rm ij}$ enter the automaton through the probabilities
\[
p_{\mathrm{eq}} = N\, dt/\tau,
\qquad
p_{\rm ij} = N f_{\rm ij}\, dt,
\]
with the diagonal term determined by normalization,
\[
p_{\rm jj}=1-\sum_{\rm i\neq j} p_{\rm ij},
\]
so that a sampled spin either undergoes local equilibration, attempts migration to a neighboring reservoir, or remains unchanged.

At each elementary update, a spin is first chosen uniformly from the full population. With probability $p_{\mathrm{eq}}$, the spin is updated according to the local Ising dynamics in its current reservoir, using the reservoir magnetization to evaluate the corresponding energy change and applying the usual thermal acceptance rule at inverse temperature $\beta$. Otherwise, a destination reservoir is sampled from the migration probabilities associated with the spin's reservoir of origin. Migration is only allowed for spins in the up state; if the selected spin is down, the attempted transfer is rejected. When an up spin migrates, the occupation numbers and magnetizations of the origin and destination reservoirs are updated accordingly, while the spin retains its subpopulation identity.

In this way, the automaton provides a microscopic realization of the coupled migration--relaxation process: local ordering is driven by Ising equilibration within reservoirs, while inter-reservoir coupling arises from directed stochastic fluxes of up spins at rates set by $f_{\rm ij}$.

\begin{figure}
	\includegraphics[width=\linewidth]{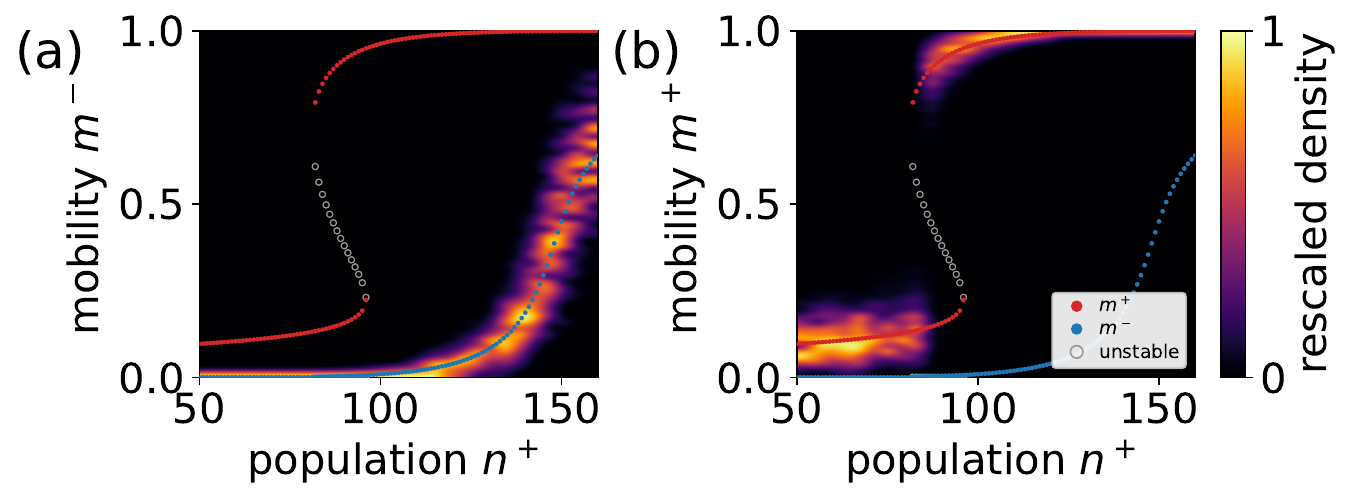}
	\includegraphics[width=\linewidth]{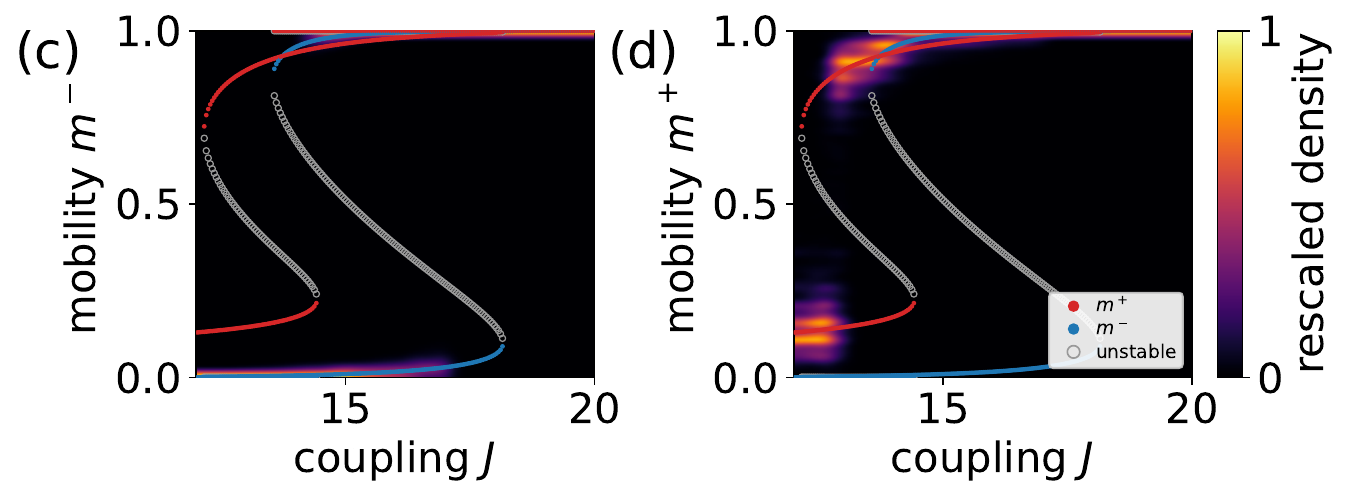}
	\caption{Comparison of mean-field solutions for average mobility for mobile and immobile populations $m^+$ and $m^-$, respectively, with automaton simulations. For visual clarity, density is normalized such that the maximum value is 1 for a given value of $n^+$ or $J$.}
\end{figure}

\begin{figure}
\centering
\includegraphics[width=.59\linewidth]{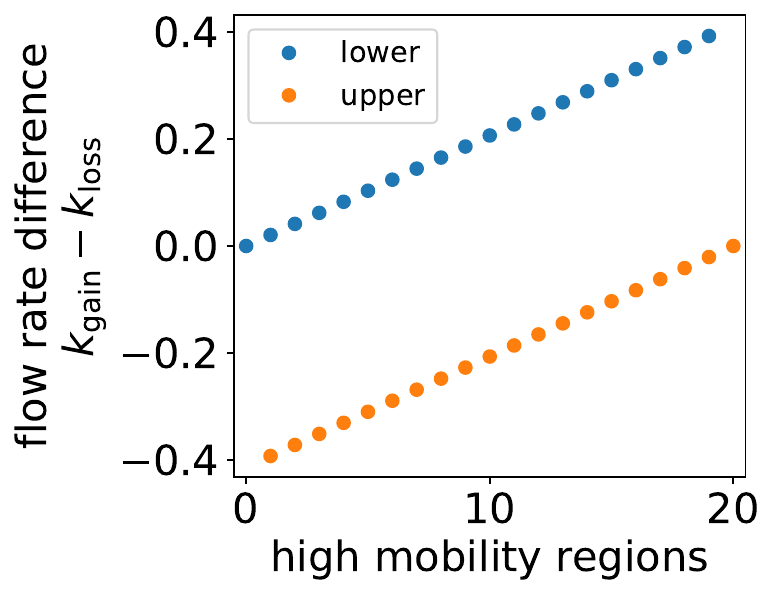}
\caption{Net migrant-flow rate driving a single region as a function of the number of regions already on the high-mobility branch, in a fully connected network of $r=20$. We plot the difference between the gain (inflow) and loss (outflow) rates of mobile ($+$) migrants, $k_{\rm gain}-k_{\rm loss}$, i.e.~the net drift of $n^+$, for a region on the lower (blue) and upper (orange) branch of the mobility solution manifold. As more regions occupy the high-mobility branch, a lower-branch region acquires positive drift (destabilizing the low-mobility configuration) while an upper-branch region's drift rises toward zero (stabilizing the high-mobility configuration); the zero crossing marks the collective tipping point discussed in the main text. Evaluated at the center of the fold bifurcation with $h^+=-2.52$, $h^-=-10.52$, $J=12$, $n^+=89$, $n=200$.}
\label{fig:drift_transition}
\end{figure}

\section{Stability of continuous transition for two regions}\label{sec:C_transition}
Let us consider the dynamics of a single continuous phase transition in one of two regions. As stated in the main text, the number of $\pm$~migrants in region~a evolves according to
\begin{equation}
  \dot{n}_{\rm a}^\pm
  = -\,n_{\rm a}^\pm\, m^\pm\!(n_{\rm a}^\pm)
    + (n^\pm - n_{\rm a}^\pm)\, m^\pm\!(n^\pm - n_{\rm a}^\pm)\,,
  \label{eq:SI_dynamics}
\end{equation}
having set the migration rate $f=1$, which only sets the relevant timescale. The total number of migrants $n^\pm = n_{\rm a}^\pm + n_{\rm b}^\pm$ is conserved for each of the ($+$) and
($-$) populations, and $m^\pm(\cdot)$ denotes the mean-field magnetization, which is the same function for both regions by symmetry---the symmetric configuration $n_{\rm a}^\pm = n_{\rm b}^\pm = n^\pm/2$ is always a fixed point.

We analyze the stability of Eq.~\eqref{eq:SI_dynamics} under a
small perturbation~$\epsilon$ away from this fixed point,
writing $n_{\rm a} + \epsilon$ and $n_{\rm b} - \epsilon$
(suppressing the $\pm$ superscript except where ambiguous).  The
exchange symmetry ${\rm a}\leftrightarrow{\rm b}$
($\epsilon\to-\epsilon$) ensures that all even-order terms
vanish in the Taylor expansion around the fixed point, yielding
to leading order

\begin{equation}
  \dot\epsilon
  = \alpha_1\,\epsilon + \alpha_3\,\epsilon^{3}
  + \mathcal{O}(\epsilon^{5})\,.
  \label{eq:SI_expansion}
\end{equation}
The coefficients are

\begin{align}
  \alpha_1 &= -2\,m - 2n_{\rm a}\,m'\,,
    \label{eq:SI_alpha1} \\[4pt]
  \alpha_3 &= -m'' - \frac{n_{\rm a}}{6}\,m^{(3)}\,,
    \label{eq:SI_alpha3}
\end{align}

where all terms are evaluated at the symmetric fixed point $n_{\rm a}^\pm = n^\pm\!/2$.  The total derivatives $m'\equiv dm^\pm/dn_{\rm a}^\pm$, $m''$, and~$m^{(3)}$ (with respect to the region population $n_{\rm a}^\pm$ along the conserved cut $n_{\rm a}^\pm+n_{\rm b}^\pm=n^\pm$) are obtained from the self-consistency relation, Eq.~\eqref{eq:self-consistency}, which also implies that the $\epsilon^2$ term is 0. Note that the susceptibility alone is not sufficient to describe the stability of the solution because the number of migrants changes under the flow dynamics.

The linear coefficient $\alpha_1$ captures the competition between migrant flow that drives the perturbation $\epsilon$ back to 0 and the resulting change in magnetization that either reinforces outflow (for ($-$) migrants) or dampens it (for ($+$) migrants). For the fixed point to be unstable, perturbations away from the fixed must grow, or $2m > -n\,m'$.  This implies that
\begin{align}
	\frac{m_{\rm a}^\pm}{n_{\rm a}^\pm} &< \left| \frac{dm_{\rm a}^\pm}{dn_{\rm a}^\pm} \right|\\
	1 &< \left| \frac{d\log m^\pm_{\rm a}}{d\log n_{\rm a}^\pm} \right|\label{eq:stability relation}
\end{align}
as the condition for symmetry breaking in the population density per region. Near marginal case, random fluctuations may lead to intermittent deviations from the deterministic solution indicated in Eq~\eqref{eq:stability relation}.




\begin{figure}\centering
	\includegraphics[width=\linewidth]{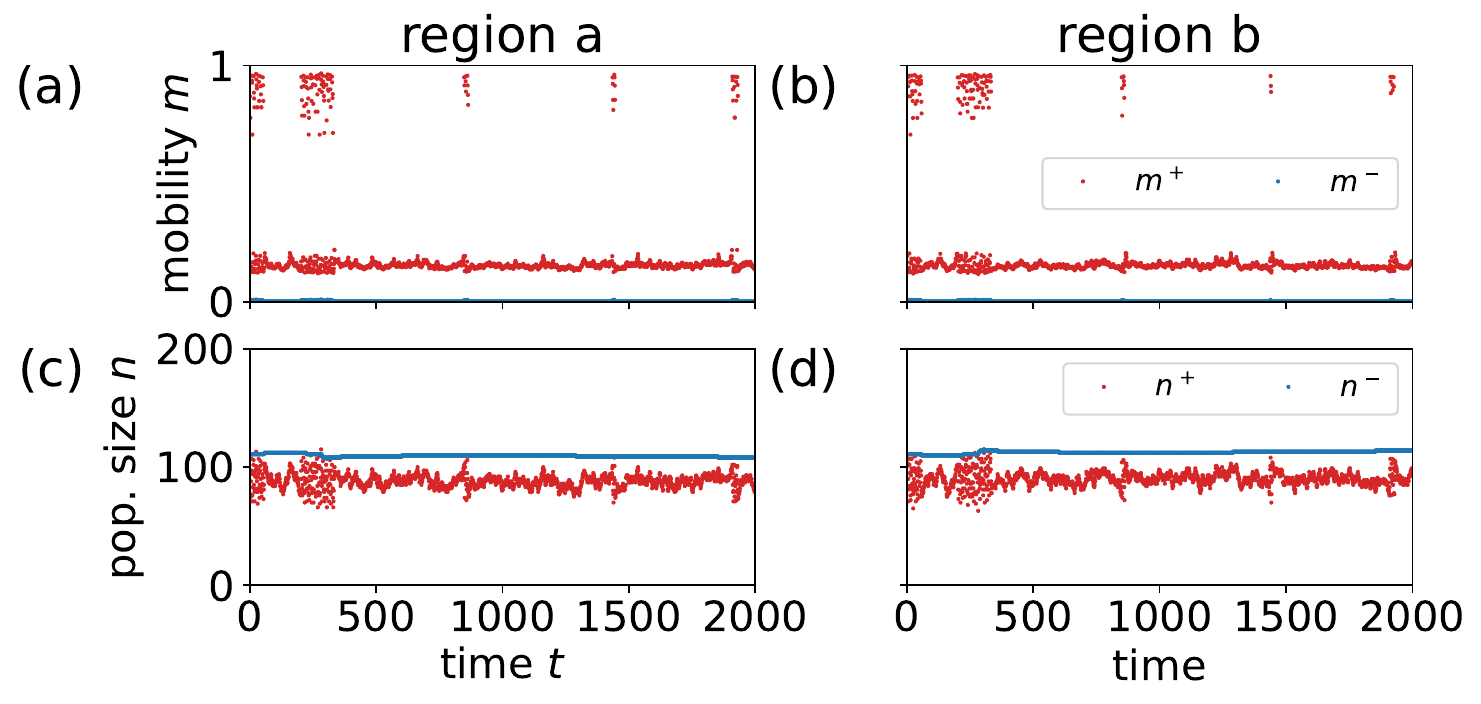}
	\caption{Desynchronization at fast $f$. When $f$ is too large, the cycle periods become short and the cycles become sensitive to the high-frequency random fluctuations that can destabilize them instead of averaging out. We show the mean mobility of the populations $m^+(t)$ and $m^-(t)$.}\label{fig:large f}
\end{figure}

\section{Data sets}
The UNHCR PRMN runs survey stations located all around Somalia that tracks the flow of migrants across various routes. For any given survey response, we have an estimate of the size of the group as well as indications of origin and destination. We can sum over all stations within a district in order to get a sense of how many people are moving from district to district---even if it is difficult to keep precise track how fast they are moving and therefore exactly when they left a region. 

The Italian Maritime Service keeps a record of the number of migrants that they register from boats intercepted in Italian waters. Given the distance that it is possible to move on these boats, they are leaving the northern coast of Africa. 

The first observation is that these statistics are not Poissonian. This may be because 1) the dynamics of group formation generate non-Poissonian distributions, 2) the distribution group sizes from different origins are highly heterogeneous, 3) the drivers fluctuate, or 4) rates at which people are leave a region fluctuate such as in our dynamics. Needless to say, we are only considering the potential impact of 4.

\section{First-passage time for the first region to cross the fold}
\label{app:fpt_crossing}
We consider the time for the first region to cross the fold bifurcation as (i)~a single region as an Ornstein--Uhlenbeck (OU) first passage as a lower bound on the crossing time, (ii)~the saddle-node correction that makes the OU estimate a lower bound, and (iii)~the large-$r$ scaling $\tau\sim1/\log r$.

We consider the case where all the regions are poised symmetrically between the two ends of a fold bifurcation such that the distance to either endpoint is $w/2$. Because a crossing region gains ($+$) migrants with little change in the ($-$) population (here $m^-\approx0$), $w$ is the fold width along the $n^-$-fixed cut. Near the lower branch the up-population $n^+$ of one region performs a biased random walk under Poissonian inflow and outflow of mobile migrants. The net per-step change is a difference of Poisson counts, or a Skellam distribution with variance $\sigma^2 = 2 f\,\bar m^+\bar n^+$. 

\begin{figure}[t]
\centering
\includegraphics[width=\linewidth]{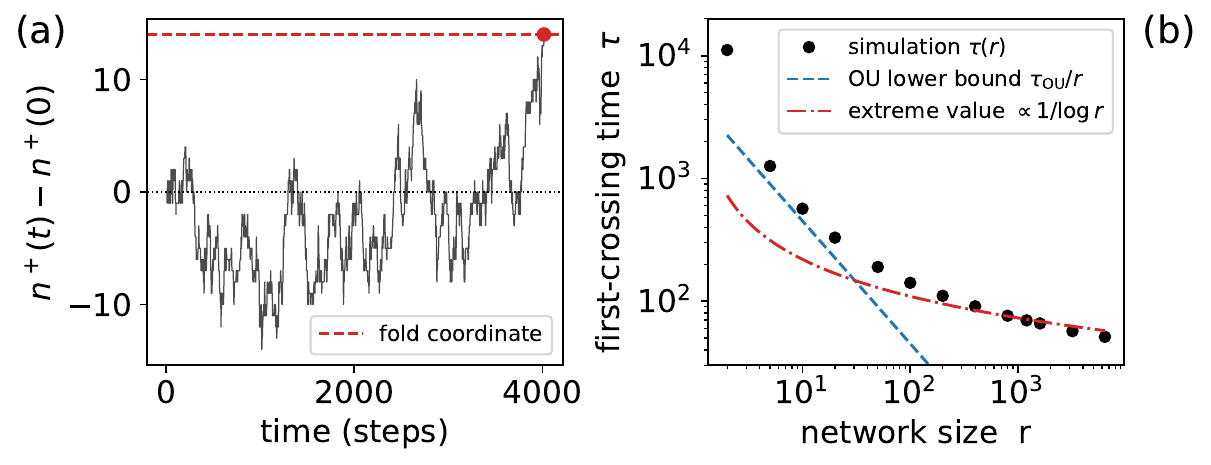}
\caption{First-passage times. (a)~A representative single-region trajectory. Given low mobility in ($-$) migrants ($n^-\approx 0$), the displacement $n^+(t)-n^+(0)$ wanders in the restoring well before crossing the fold---where the lower branch vanishes along this $n^-$-fixed path (dashed line)---on a rare sustained excursion (red point). (b)~Mean first-crossing time $\tau$ versus network size $r$ (points). We compare the single-region OU bound $\tau_{\rm OU}/r$ (dashed) against the extreme-value law $\propto1/\log r$ (dash-dot). Automaton simulation is shown as black points. Parameters $h^+=-2.52$, $h^-=-10.52$, $J=12$, $n=200$, start $n^+=89$.}
\label{fig:fpt}
\end{figure}

The inflow from other regions serves as as a linear restoring drift with ``spring constant'' $k = f\big[\,m^+ + n^+\,\partial m^+/\partial n^+\big|_{n^-}\big]$. Taking the linear approximation, one region is an OU walker, and the mean first-passage time comes from the Siegert integral $\tau_{\rm OU}=(\sqrt\pi/k)\int_0^{u_a}e^{u^2}[1+\mathrm{erf}(u)]\,du$, with $u_a=w\sqrt{k/8D}$ and $D=\sigma^2/2$. This means that a confined region wanders mostly below the start and crosses only on a rare sustained excursion (Fig.~\ref{fig:fpt}a).

The locally linear approximation uses the branch slope $\partial m^+/\partial n^+$ calculated at the symmetric starting point, but the lower branch steepens toward the fold because of the saddle-node geometry, a restoring force that grows nonlinearly along the path. If we now integrate the restoring work $\int_0^{w/2} k(\delta)\,\delta\,d\delta$, we obtain a substantially larger expected crossing time because the crossing time is exponential in the barrier. In the test case ($r=20$; $h^+=-2.52$, $h^-=-10.52$, $J=12$, $n=200$, start $n^+=89$, giving $\sigma^2\approx0.27$ and $k\approx0.0034$) the directly measured single-region crossing time, $\tau \approx6\times10^3$ steps compared to the linearized OU estimate $\tau_{\rm OU} \approx4.5\times10^3$.

Numerical arguments show that the network's first jump displays two regimes depending on whether we have a few regions (Poisson-rare, $\tau\sim\tau_{\rm single}/r$, when each region escapes as a slow memoryless event) or many regions (extreme-value, $\tau\sim w^2/8\sigma^2\log r$, when spontaneous alignment of migrants drives a fast diffusive fluctuation before the restoring force can act). This is because conservation of migrant number effectively matters less for the dynamics as we consider larger $r$, and we see a gradual crossover across the range $r=2$ to $r=6400$ (Fig.~\ref{fig:fpt}b), with slow convergence to the extreme-value tail. 

\begin{figure}[t]
\centering
\includegraphics[width=\linewidth]{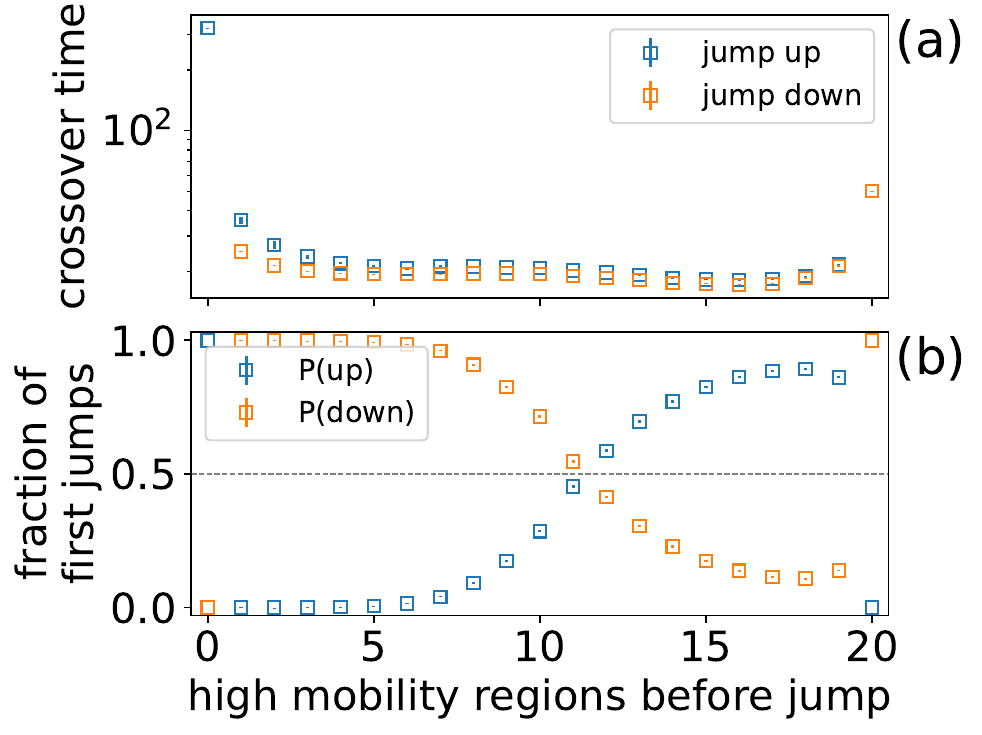}
\caption{Crossover time and direction of the next region to cross the fold, in a fully connected network of $r=20$, as a function of the number of regions already on the high-mobility branch. (a)~Mean first-crossing time, separated into an additional region switching up to the high-mobility branch (blue) and one of the high-mobility regions relaxing back down (orange); points are stochastic mean-field simulations with standard errors, conditioned on a single-region crossing. The two timescales converge more regions join the high-mobility configuration. (b)~Fraction of first crossings in each direction (up blue, down orange); the dashed line marks equal odds, crossed near twelve of the twenty regions, beyond which switching a further region up is more likely (cf.~Fig.~\ref{fig:drift_transition}). Parameters $h^+=-2.52$, $h^-=-10.52$, $J=12$, $n=200$, start $n^+=89$.}
\label{fig:crossover_split}
\end{figure}

Away from the symmetric start, both the crossing time and its likely direction depend on how many regions already occupy the high-mobility branch (Fig.~\ref{fig:crossover_split}).

\section{Estimated oscillation period for \textit{r=2}}
We estimate the period for an oscillation in the dyadic $r=2$ network. Assuming that the mobility on the upper and lower branches of the solution manifold vary weakly with $n^+$, which also implies that $n^-$ remains almost constant, the typical time comes from
\begin{align}
	\dot n^+(t) &= -f \left[ m^+_{\rm u} n^+(t) - m^+_{\rm l} (n-n^+(t)) \right],
\end{align}
using the subscripts u and l to denote the upper and lower branches of the solution. Given width of the fold bifurcation $w$ and the upper end of the fold $n^+_0$, the period is
\begin{align}
\begin{aligned}
	t &= \frac{2}{f(m^+_{\rm u} + m^+_{\rm l})}\left[ \log\left( n^+_0 - \frac{m^+_{\rm l} n}{m^+_{\rm u} + m^+_{\rm l}} \right) - \right.\\
	   & \left. \qquad\log\left(n^+_0 + w - \frac{m^+_{\rm l} n}{m^+_{\rm u} + m^+_{\rm l}} \right) \right].
\end{aligned}
\end{align}

The calculation is more complicated for a cycle with $r>2$ because the width of the traveling pulse can be larger than one region.

%

\section{Solution boundaries}\label{sec:sol_boundary}
The self-consistent magnetization obeys
\begin{align}
    m^{+}(m) &= \frac{e^{h^{+}+Jm}}{1+e^{h^{+}+Jm}}, \label{eq:sm_mp}\\
    m^{-}(m) &= \frac{e^{h^{-}+Jm}}{1+e^{h^{-}+Jm}}, \label{eq:sm_mm}\\
    m &= \frac{n^{+}}{n}\,m^{+}(m) + \frac{n-n^{+}}{n}\,m^{-}(m) \;\equiv\; G(m),
    \label{eq:sm_G}
\end{align}
where we have dropped the $-Jm^{\pm}/n$ self-interaction correction of Eq~\eqref{eq:self-consistency}, negligible for the large populations $n$ considered here. A fixed point of the dynamics is
a solution of $m = G(m)$, i.e.\ an intersection of $G(m)$ with the diagonal
$m=m$. The system transitions between one and two stable solutions through a
fold bifurcation, which occurs when $G(m)$ becomes tangent to the
diagonal. This pair of conditions reads
\begin{equation}
    G(m) - m = 0
    \qquad \text{and} \qquad
    \frac{\partial G(m)}{\partial m} = 1,
    \label{eq:sm_foldcond}
\end{equation}
where the tangency condition can be written explicitly using
$\partial m^{\pm}/\partial m = J\,m^{\pm}(1-m^{\pm})$. 

Now, we observe that varying $h^+$ and $h^-$ translates the $m^{+}$ and
$m^{-}$ sigmoid curves, \eqref{eq:sm_mp}-\eqref{eq:sm_mm}, along the $m$ axis without
altering their shape. When the field gap $\Delta h = h^{+} - h^{-}$ is large (as
it is for our parameters), the two sigmoids are well separated, so for any $m$
one of the two components is effectively saturated:
\begin{equation}
    m^{-}(m) \approx 0
    \qquad \text{or} \qquad
    m^{+}(m) \approx 1.
    \label{eq:sm_approx}
\end{equation}
(The intermediate regime with $m^{-}(m)\approx 0$ \emph{and}
$m^{+}(m)\approx 1$ yields only the trivial solution and is therefore not
considered here.) Each case in \eqref{eq:sm_approx} isolates one population and
produces one family of fold boundaries, which we take in turn.

In the first case, $m^{-}(m) \approx 0$, \eqref{eq:sm_G} reduces to
$m \simeq (n^{+}/n)\,m^{+}$, so that
\begin{equation}
    m^{+} \simeq \frac{e^{h^{+}+Jn^{+}m^{+}/n}}{1+e^{h^{+}+Jn^{+}m^{+}/n}}.
\end{equation}
Inverting the sigmoid gives
\begin{equation}
    h^{+} \simeq \ln\!\frac{m^{+}}{1-m^{+}} - \frac{J n^{+} m^{+}}{n},
    \label{eq:sm_hp_implicit}
\end{equation}
while the tangency condition $\partial G/\partial m = 1$ gives us,
\begin{equation}
    \frac{J n^{+}}{n} \simeq \frac{1}{m^{+}(1-m^{+})}.
    \label{eq:sm_tangent_mp}
\end{equation}
Substituting \eqref{eq:sm_tangent_mp} into \eqref{eq:sm_hp_implicit} eliminates
the explicit dependence on $J n^{+}$ and yields a relation between the field and
the magnetization at the bifurcation,
\begin{equation}
    h^{+} \simeq \ln\!\frac{m^{+\ast}}{1-m^{+\ast}}
    - \frac{1}{1-m^{+\ast}},
    \label{eq:sm_hp_star}
\end{equation}
where $m^{+\ast}$ denotes the magnetization values that satisfy the fold
conditions \eqref{eq:sm_foldcond}. For a fixed $h^{+}$, equation
\eqref{eq:sm_hp_star} admits two roots $m^{+\ast}$, corresponding to the two fold
bifurcations in $m^{+}$. Inserting each root into \eqref{eq:sm_tangent_mp} (with
$n = 200$ and $h^{+} = -2.52$) gives the two critical products

\begin{equation}
\label{eq:4a_plus}
    J n^{+} \simeq 1141
    \qquad \text{and} \qquad
    J n^{+} \simeq 962 ,
\end{equation}

which trace two of the solution boundaries in Figure \ref{fig:solution landscape}a.

In the second case, $m^{+}(m) \approx 1$, \eqref{eq:sm_G} reduces to
$m \simeq n^{+}/n + m^{-} (n-n^{+})/n$, so that
\begin{equation}
    m^{-} \simeq \frac{e^{h^{-}+J[n^{+}+(n-n^{+})m^{-}]/n}}{1+e^{h^{-}+J[n^{+}+(n-n^{+})m^{-}]/n}}.
\end{equation}
Inverting the sigmoid gives
\begin{equation}
    \frac{J n^{+}}{n} + \frac{J(n-n^{+})\,m^{-}}{n}
    \simeq \ln\!\frac{m^{-}}{1-m^{-}} - h^{-},
    \label{eq:sm_mm_implicit}
\end{equation}
and the tangency condition $\partial G/\partial m = 1$ gives us,
\begin{equation}
    \frac{J(n-n^{+})}{n} \simeq \frac{1}{m^{-}(1-m^{-})}.
    \label{eq:sm_tangent_mm}
\end{equation}
Combining \eqref{eq:sm_mm_implicit} and \eqref{eq:sm_tangent_mm} and solving for
the coupling and the population size yields a parametric form of the boundary in
terms of the bifurcation magnetization $m^{-\ast}$,

\begin{equation}
\label{eq:4a_minus}
\left.
\begin{aligned}
    J &= \frac{1}{m^{-\ast}(1-m^{-\ast})}
    + \ln\!\frac{m^{-\ast}}{1-m^{-\ast}}
    - \frac{1}{1-m^{-\ast}} - h^{-}, \\
    n^{+} &= \frac{n}{J}
    \left\{\ln\!\frac{m^{-\ast}}{1-m^{-\ast}}
    - \frac{1}{1-m^{-\ast}} - h^{-}\right\},
\end{aligned}
\right\}
\end{equation}

where $m^{-\ast}$ are the magnetization values satisfying
\eqref{eq:sm_foldcond}. Sweeping $m^{-\ast}$ over $(0,1)$ traces the remaining
fold boundary in the $(n^{+}, J)$ plane. Together, \eqref{eq:4a_plus} and
\eqref{eq:4a_minus} produce the analytical solution boundaries (black curves) in
Figure \ref{fig:solution landscape}a, in agreement with the numerically
determined number of stable solutions.

We now turn to panel (b), the swept field $h^{+}$ enters only through $m^{+}$, whereas
$m^{-} = e^{h^{-}+Jm^{\ast}}/(1+e^{h^{-}+Jm^{\ast}})$ depends only on fixed parameters and is
therefore a known number once the bifurcation magnetization $m^{\ast}$ is
chosen. For panel (a), by contrast, the swept axis $J$ enters both $m^{+}$ and
$m^{-}$ through the product $Jm$, so no analogous elimination is possible and
the bifurcation magnetization must be found by a numerical root solve, hence we needed the approximation there.

The two fold conditions \eqref{eq:sm_foldcond} at a bifurcation magnetization
$m^{\ast}$ read
\begin{align}
    \frac{n^{+}}{n}\,m^{+} + \frac{n-n^{+}}{n}\,m^{-} &= m^{\ast},
    \label{eq:sm_b_G}\\
    \frac{n^{+}}{n}\,m^{+}(1-m^{+})
    + \frac{n-n^{+}}{n}\,m^{-}(1-m^{-}) &= \frac{1}{J},
    \label{eq:sm_b_tangent}
\end{align}
where $m^{+}, m^{-}$ are shorthand for $m^{+}(m^{\ast}), m^{-}(m^{\ast})$ from
\eqref{eq:sm_mp}-\eqref{eq:sm_mm}. We march the boundary in $m^{\ast}$, treating
$m^{-}$ as known and $(n^{+}, m^{+})$ as the unknowns. Each condition
\eqref{eq:sm_b_G}-\eqref{eq:sm_b_tangent} solves linearly for $n^{+}/n$,
\begin{equation}
\begin{aligned}
\frac{n^{+}}{n}
&=
\frac{m^{\ast}-m^{-}}{m^{+}-m^{-}},
\\[0.5em]
\frac{n^{+}}{n}
&=
\frac{1/J - m^{-}(1-m^{-})}
     {m^{+}(1-m^{+}) - m^{-}(1-m^{-})}.
\end{aligned}
\label{eq:sm_b_npfrac}
\end{equation}
Equating the two expressions and factoring the right-hand denominator,
$m^{+}(1-m^{+}) - m^{-}(1-m^{-}) = (m^{+}-m^{-})(1-m^{+}-m^{-})$, the common
factor $(m^{+}-m^{-})$ cancels and leaves a relation \begin{equation}
\begin{split}
(m^{\ast}-m^{-})(1-m^{+}-m^{-})
= \frac{1}{J} - m^{-}(1-m^{-}) \\
\Longrightarrow\quad
m^{+}
= 1 - m^{-}
- \frac{1/J - m^{-}(1-m^{-})}{m^{\ast}-m^{-}}.
\end{split}
\label{eq:sm_b_mp}
\end{equation}
The field is recovered at the end by inverting the sigmoid, $h^{+} =
\ln\!\frac{m^{+}}{1-m^{+}} - Jm^{\ast}$. Collecting these results, the panel (b)
boundary is the exact parametric curve, for $m^{\ast} \in (0,1)$,

\begin{equation}
\label{eq:4b}
\left.
\begin{aligned}
    m^{-} &= \frac{e^{h^{-}+Jm^{\ast}}}{1+e^{h^{-}+Jm^{\ast}}}, \\
    m^{+} &= 1 - m^{-}
    - \frac{1/J - m^{-}(1-m^{-})}{m^{\ast}-m^{-}}, \\
    n^{+} &= n\,\frac{m^{\ast}-m^{-}}{m^{+}-m^{-}},
    \qquad
    h^{+} = \ln\!\frac{m^{+}}{1-m^{+}} - Jm^{\ast},
\end{aligned}
\right\}
\end{equation}

retaining the points with $m^{+} \in (0,1)$ and $n^{+} \in [0,n]$. Sweeping
$m^{\ast}$ over $(0,1)$ traces the locus $\{(n^{+}, h^{+})\}$ given by
\eqref{eq:4b}, the analytical solution boundary (black curve) in Figure
\ref{fig:solution landscape}b, again in agreement with the numerically
determined number of stable solutions.

\section{Necessary conditions for fold bifurcation in (+) population}\label{sec:min_hp}
We consider the fold conditions for the ($+$) population in the large field-gap limit ($\Delta h = h^{+} - h^{-}$ large), where Equation \eqref{eq:sm_hp_star} relates the field $h^+$ to the bifurcation magnetization $m^+$, and the accompanying tangency condition, Equation \eqref{eq:sm_tangent_mp}, fixes the product $Jn^+/n$. Two necessary conditions follow.

First, the tangency condition reads
\begin{equation}
    \frac{Jn^+}{n} \simeq \frac{1}{m^+(1-m^+)}.
\end{equation}
The right-hand side is minimized at $m^+ = 1/2$, where it equals $4$. Hence a fold can occur only if
\begin{equation}\label{eq:sm_Jcond}
    \frac{Jn^+}{n} \geq 4,
\end{equation}
since for $Jn^+/n < 4$ the tangency condition has no solution $m^+ \in (0,1)$ and the self-consistency map cannot become tangent to the diagonal for any value of $h^+$.

\begin{figure}[t]
\centering
\includegraphics[width=\linewidth]{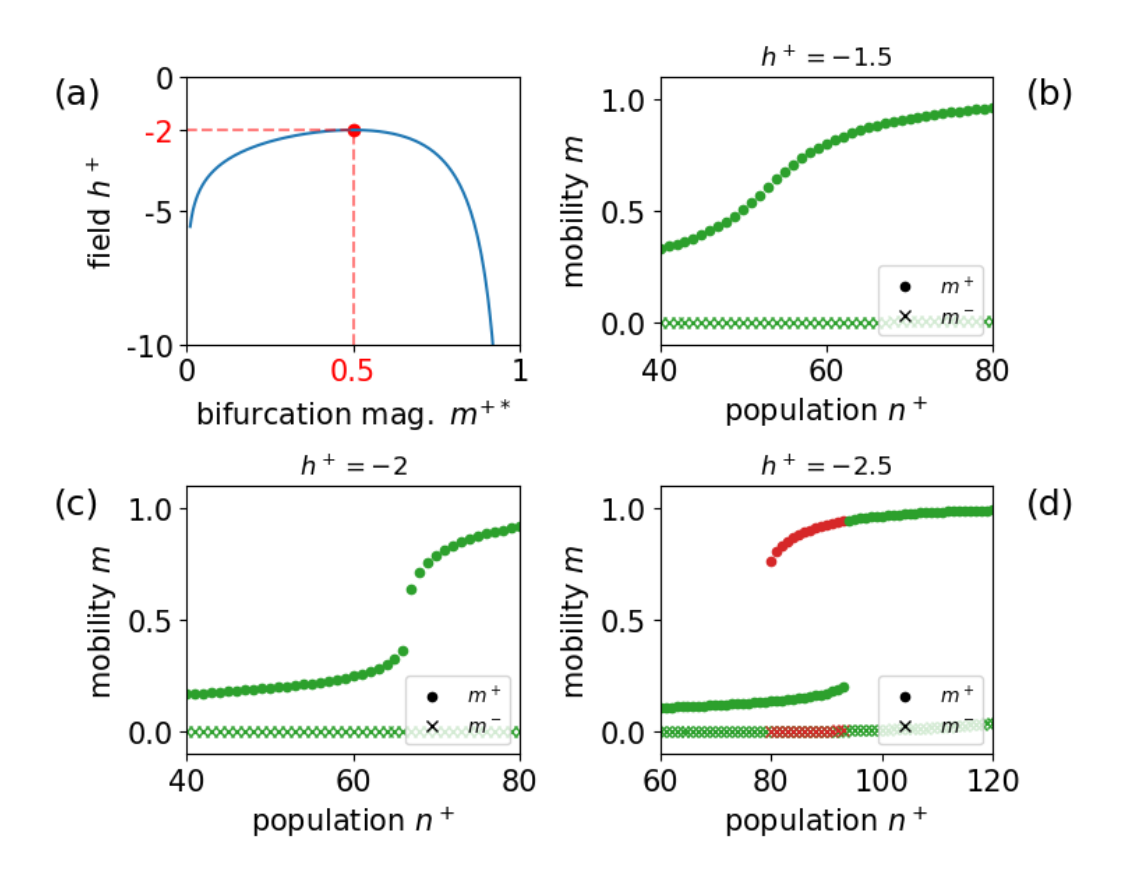}
\caption{Necessary condition for a fold bifurcation in the $(+)$ population. (a) The relationship between the field $h^+$ and the bifurcation magnetization $m^{+*}$ defined by Eq.~(\eqref{eq:sm_hp_star}); shows that the maximum $h^+$ admitting a fold is $h^+ = -2$. (b--d) Mobility solutions $m^+$ and $m^-$ versus $n^+$ as $h^+$ is varied, confirming that the $(+)$ population develops a fold bifurcation only once $h^+$ drops below $-2$. In these panels, green and red distinguish the solutions: green-only regions have a single solution, while regions with red points have two. Parameters are $h^-=-10.52$ and $J=12$}
\label{fig:vary_hp_SI}
\end{figure}

Second, eliminating $Jn^+$ between the tangency and self-consistency conditions yields Equation \eqref{eq:sm_hp_star}, which defines the relationship between $h^+$ and $m^+$ at which the ($+$) population undergoes a fold bifurcation and follows the fold condition Equation \eqref{eq:sm_foldcond}. By sweeping $m^+$ across its domain ($0 \leq m^+ \leq 1$), we find that the largest value of $h^+$ that satisfies Equation \eqref{eq:sm_hp_star} is $-2$, as shown in Figure \ref{fig:vary_hp_SI}a, so that
\begin{equation}\label{eq:sm_hcond}
    h^+ < -2
\end{equation}
is a necessary condition for the fold.

This result is confirmed by examining the evolution of the solutions as $h^+$ is varied (Figure \ref{fig:vary_hp_SI}b,c,d). Taken together, a fold bifurcation in the ($+$) population requires both $h^+ < -2$ and $Jn^+/n \geq 4$.

\section{Examples of migration dynamics from Somalia}\label{sec:data_dynamics}
\begin{figure}[t!]
\centering
\includegraphics[width=\linewidth]{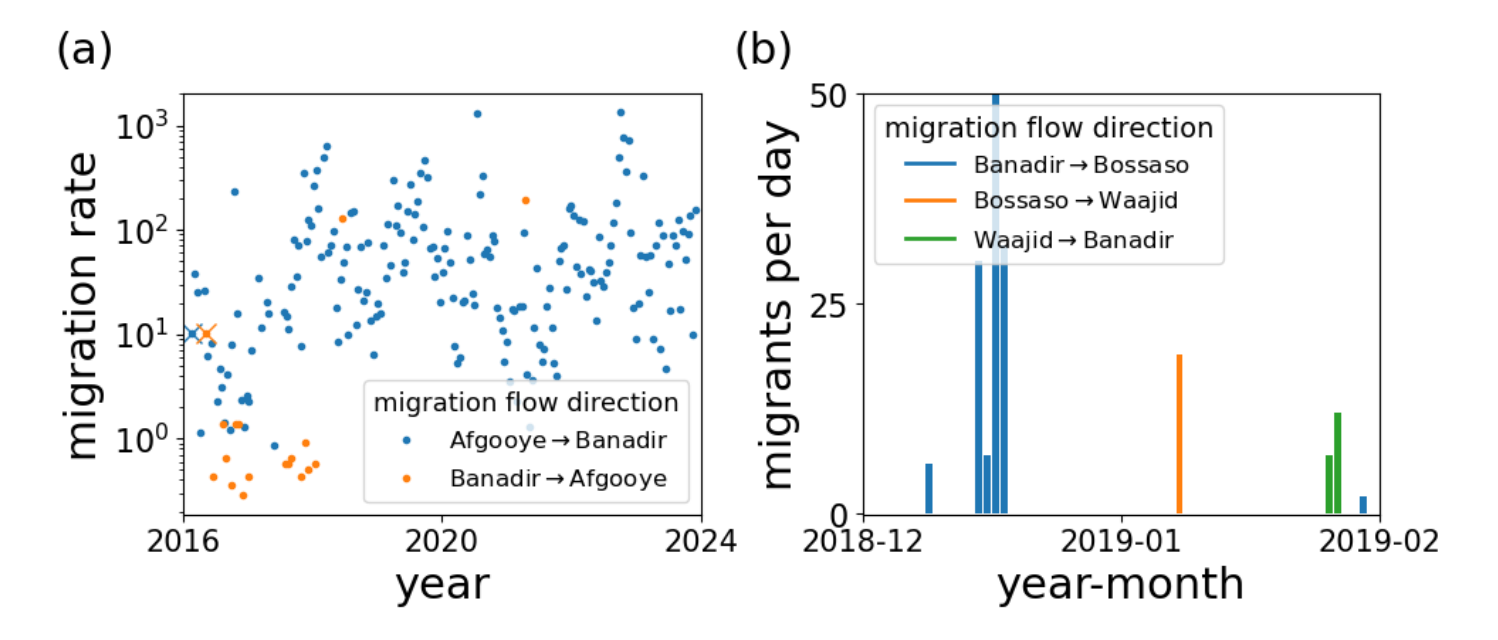}
\caption{Examples of migration dynamics from high-resolution data. a) Migration flows between two regions in Somalia shows asymmetric rates, analogous to the symmetry breaking we show in the model. Rates are calculated using 14 day time windows. Cross markers (×) denote the first measurement in each direction, showing that migration rates increase over time in one direction but decay to zero in the other. y-axis is shown in log scale. b) Traveling migration pulse in three region cycle induced by migration from region a to region b (blue). Data sparsity leaves few such examples to consider.}
\label{fig:som_data}
\end{figure}

Recent advances in the availability of fine-grained migration data have enabled detailed analyses of real-world migration dynamics. However, such high-resolution datasets remain openly accessible for only a handful of regions worldwide. Here, we examine migration data from Somalia, where interregional migration flows are recorded at daily temporal resolution. Consistent with our theoretical predictions, we observe the emergence of source-sink region pairs and cyclic cascade patterns, as illustrated in Figure~\ref{fig:som_data}.

Source-sink dynamics are observed across multiple regional pairs in Somalia. Figure \ref{fig:som_data}a presents one such example involving Banadir, the country's main administrative region, and Afgooye in the Lower Shebelle region. The migration rate over time clearly indicates that Banadir functions as a sink, a point made in the social and political science literature \cite{blair2026dynamics,adam2025community}.

Next, we consider a triadic regional network based on migration routes that have been identified in the Somalian migration data \cite{oh2024emergent}. Focusing on cycles, we examine a representative example consisting of the regions of Banadir, Bossaso and Waajid. As shown in Figure~\ref{fig:som_data}b, this triad exhibits a traveling pulse in the time period we consider.

Despite the high temporal resolution of the dataset, the analysis remains constrained by data sparsity. While meaningful patterns can be reliably identified at the pairwise and triadic levels, extending the analysis to higher-order migration structures proves challenging.

\end{document}